\documentclass[11pt]{article}

\usepackage[english]{babel}
\usepackage[latin1]{inputenc}
\usepackage[T1]{fontenc}

\usepackage{amsmath}
\usepackage{graphicx}
\usepackage[colorinlistoftodos]{todonotes}
\usepackage{xcolor}
\usepackage{pgfplots}
\usepackage{bm}
\usepackage{todonotes}
\usepackage{multirow}
\usepackage{arydshln}

\usepackage{algpseudocode}
\usepackage{algorithm}

\usepackage{setspace}

\usepackage{bm}
\usepackage{upgreek}

\usepackage{amsmath}
\usepackage{physics}

\usepackage{natbib}

\usepackage[noblocks]{authblk}
\usepackage[margin=1in]{geometry}
\usepackage[title]{appendix}

  \DeclareTextFontCommand\textsfi{\usefont{OT1}{cmss}{m}{sl}}
  \DeclareMathAlphabet\mathsfi            {OT1}{cmss}{m}{sl}
  \DeclareTextFontCommand\textsfb{\usefont{OT1}{cmss}{bx}{n}}
  \DeclareMathAlphabet\mathsfb            {OT1}{cmss}{bx}{n}
  \DeclareTextFontCommand\textsfbi{\usefont{OT1}{cmss}{m}{sl}}
  \DeclareMathAlphabet\mathsfbi            {OT1}{cmss}{m}{sl}
  \IfFileExists{t1phv.fd}
  {\DeclareTextFontCommand\textsfbi{\usefont{T1}{phv}{b}{it}}
  \DeclareMathAlphabet\mathsfbi            {T1}{phv}{b}{it}
  }{}
  \IfFileExists{ot1phv.fd}
  {\DeclareTextFontCommand\textsfbi{\usefont{OT1}{phv}{b}{it}}
  \DeclareMathAlphabet\mathsfbi            {OT1}{phv}{b}{it}
  }{}

\newcommand\affiliation[1]{\gdef\@affiliation{\let\aff\aff@inst#1}}
\gdef\@affiliation{}

\def\aff#1{\ignorespaces\textsuperscript{#1}}

\numberwithin{equation}{section}

\renewenvironment{abstract}
{\begin{quote}
\noindent \rule{\linewidth}{.5pt}\par{\bfseries \abstractname.}}
{\medskip\noindent \rule{\linewidth}{.5pt}
\end{quote}
}

\newcommand{\vecd}[1]{\mathbf{#1}} 

\newcommand{\q}{\mathbf{q}}
\newcommand{\qm}{\mathbf{\bar{q}}}
\newcommand{\qf}{\mathbf{q}'}

\newcommand{\tc}[1]{\mathsfbi{#1}}	

\newcommand{\partderiv}[2]{\frac{\partial #1}{\partial #2}}

\usepackage{hyperref}

\definecolor{darkblue}{rgb}{0,0,0.80}

\hypersetup{
    colorlinks=true,
    linkcolor={darkblue},
    citecolor={darkblue},
    filecolor=black,
    urlcolor={darkblue},
    plainpages=false,
}

\makeatletter
\renewcommand*\env@matrix[1][\arraystretch]{%
  \edef\arraystretch{#1}%
  \hskip -\arraycolsep
  \let\@ifnextchar\new@ifnextchar
  \array{*\c@MaxMatrixCols c}}
\makeatother

\title{\bf Efficient global resolvent analysis via the one-way Navier-Stokes equations. Part 2. Optimal response}

\author[1]{\bf Georgios Rigas}
\author[2]{\bf Omar Kamal}
\author[3]{\bf Aaron Towne}
\author[2]{\bf Tim Colonius}

\affil[1]{\normalsize Department of Aeronautics, Imperial College London, London SW7 2AZ, UK  }
\affil[2]{\normalsize Mechanical Engineering, California Institute of Technology, Pasadena, CA USA}
\affil[3]{\normalsize Department of Mechanical Engineering, University of Michigan, Ann Arbor, MI, USA \vspace{-1cm}}

\date{}
\begin{document}

    \maketitle
    
    \begin{abstract}
In this study, we develop an efficient approach for approximating resolvent modes via spatial marching.  Building on the methodology from Part 1, we leverage the ability of the projection-based formulation of the one-way Navier-Stokes equations (OWNS) to efficiently and accurately approximate the downstream response of the linearized Navier-Stokes equations to forcing for problems containing a slowly varying direction.  Using an adjoint-based optimization framework, forcings that optimally excite a response in the flow are computed by marching the forward and adjoint OWNS equations in the downstream and upstream directions, respectively.  This avoids the need to solve direct and adjoint globally-discretized equations, therefore bypassing the main computational bottleneck of a typical global resolvent calculation.  The method is demonstrated for a supersonic turbulent jet at Mach 1.5 and a transitional  zero-pressure-gradient flat-plate boundary layer flow at Mach 4.5, and the optimal OWNS results are validated against corresponding global calculations. \\
    \end{abstract}
    


\section{Introduction}

Fast computation of linear perturbations to two- and three-dimensional steady base flows is a common requirement for modern stability analyses, flow control efforts, and modeling of turbulence structures \citep{taira2017modal}.  In particular, resolvent (input-output) analyses in the frequency domain inform, for laminar base flows, the forcings and responses in worst-case transition scenarios \citep{monokrousos2010global}, whereas for turbulent mean flows, they are a useful way to characterize linear amplification mechanisms leading to coherent structures \citep{McKeon2010critical, Towne2018spectral}.  In both cases, one seeks to maximize a prescribed scalar quantity (norm) that represents the gain between the worst-case inputs and outputs.  While there are variations in the approaches, the typical computation involves (i) discretization of the continuous flow problem to obtain a finite-dimensional one, (ii) representing the linearized relationship between inputs and outputs as a matrix (the resolvent operator), and (iii) finding one or more singular values/vectors of the resolvent matrix via the singular value deposition (SVD).  

Apart from iterative approaches that solve analogous problems in the time domain \citep{monokrousos2010global, Martini2021efficient, Farghadan2021randomized}, there are a variety of computational issues associated with efficiently finding the singular vectors. Even when efficient randomized SVD algorithms are employed \citep{ribeiro2020randomized}, the computational expense of this procedure is dominated by solution (inversion) of the linear system.  In order to reach 2D and 3D base flows, which are typically represented by $10^5$ or more degrees of freedom, this is only feasible if the discretization produces a sparse operator, such that fast methods--typically direct multifrontal LU decomposition--can be employed in solving the system.   An additional complication that arises is that both the resolvent system and its transpose (adjoint) need to be solved (many times) such that ideally the transpose system should be solved with the same high efficiency as the primal one. Even in the sparse case, computation rate and memory bottlenecks limit the grid sizes of such solutions and/or force them to be performed in a high-performance-computing (HPC) environment.  As typically one aims to characterize the input-output relationships over a broad range of frequencies and parameters, it is of great interest to reduce the computational burden.

In Part 1 of this paper \citep{Towne2021fast}, we showed that the action of the resolvent operator on a forcing vector can be efficiently approximated via spatial marching for flows that include a slowly varying direction, i.e., one in which the mean flow is inhomogeneous but changes gradually, such as many boundary and free-shear layers.  The approximation involves parabolization of the equations, in which any disturbances traveling in the upstream direction is removed.  The resulting equations, termed the One-Way Navier-Stokes (OWNS) equations, generalize and robustify previous ad hoc approaches such as the parabolized stability equations (PSE).  


As compared to the global approach, the ``projection'' OWNS method developed in Part 1 (OWNS-P for short) achieves this solution at a FLOPS and memory cost that is lower by factors of $N_x^{a-1}/N_\beta^{a}$ and $N_x^{b}/N_\beta^{b}$, respectively.  Here $N_x$ is the number of discretization points in the streamwise (parabolized) direction, $N_\beta \ll N_x$ is a number that represents the complexity of the parabolization (see Part 1), and the scaling coefficients $1 < a < 3$ and $1 < b < 2$ estimate the FLOPS and memory cost, respectively, of the multifrontal LU.  Generally these reach their lower bounds for sparser and simply-structured matrices and the upper bounds represent dense linear algebra.  OWNS thus always provides a large memory savings and, in most scenarios, a significant speedup in computational rate as well.

In this paper, we show that the OWNS-P framework can be extended to solve, in an iterative fashion, for the singular values/vectors of the resolvent operator at the same significantly reduced computational cost and memory overhead.    In section \ref{sec:GlobalResolvent}, we formulate the global resolvent problem and provide a short review of techniques for calculating the resolvent modes. In section \ref{sec:OptimalOWNS}, we reformulate the problem of calculating resolvent modes using the OWNS-P framework, thus reducing substantially the computational cost. In section \ref{sec:results}, we demonstrate and validate the ability of the one-way Navier-Stokes (OWNS) equations to accurately reproduce the global resolvent modes from the literature for a turbulent, supersonic (Mach 1.5) jet and a laminar Mach 4.5 zero-pressure-gradient flat-plate boundary-layer.  While we focus on supersonic flows, the methodology is robust and efficient for all flow speeds.  In section \ref{sec:conclusions}, we summarize the advantages and restrictions of the optimal OWNS framework and discuss how it can be employed to compute resolvent modes in flows for which the global approach would be intractable.

\section{Global resolvent analysis}\label{sec:GlobalResolvent}

After spatial discretization in all direcctions and incorporationg of appropriate boundary conditions, the linearized compressible Navier-Stokes equations may be written in compact form as
\begin{subequations}
\begin{align}
\partderiv{ \boldsymbol{q}^{\prime} }{t} + \tc{A}_G \boldsymbol{q}^{\prime}  & =  \tc{B}_q \boldsymbol{f},
\label{eq:LNS1} \\
 \boldsymbol{y}^{\prime} & =  \tc{C}_q \boldsymbol{q}^\prime,
\label{eq:observe}
\end{align}
\label{eq:LNS}
\end{subequations}
where $\boldsymbol{q}^\prime$ is a vector of discretized dependent variables, $\boldsymbol{y}^\prime$ are the observables, $\boldsymbol{f}$ is a generic forcing term, and $\tc{B}_q$ and $\tc{C}_q$ are the prolongation and restriction operators for the forcing and observables, respectively.  The matrix $\tc{A}_G$ is a function of the base (or mean) flow about which the equations have been linearized.  If the base flow is homogenous in a coordinate direction, the solution can be decomposed into Fourier modes in that direction to reduce the number of dimensions in which the linearized equations must be discretized.  

Equation~\eqref{eq:LNS} can be expressed in the frequency domain (via the temporal Fourier transform) as
\begin{equation}\label{eq:LNS2}
    \boldsymbol{\hat{y}} = \tc{C}_q \tc{R} \tc{B}_q \boldsymbol{\hat{f}},
\end{equation}
where $\tc{R} = \left( i \omega \tc{I} + \tc{A}_G \right)^{-1}$ and $ \hat{(\cdot)} $ variables represent frequency-transformed quantities. For the remainder of the paper, we assume that all perturbed quantities are expressed in the frequency domain and drop the hat.

We now define the following global inner product
\begin{align}
\langle {\bf a},{\bf b} \rangle_G = \langle {\bf a},\tc{W}_e {\bf b} \rangle	= {\bf a}^H \tc{W}_{xyz} \tc{W}_e {\bf b} &= {\bf a}^H \tc{W}_G {\bf b}, 
\end{align}
where $H$ represents the Hermitian transpose and $\tc{W}_G$ is a positive definite weighting matrix.  $\tc{W}_G$ is constructed as $\tc{W}_G = \tc{W}_{xyz} \tc{W}_e$ where $\tc{W}_e$ is chosen so that the output represents a physical quantity of interest (e.g. energy) and $\tc{W}_{xyz}$ is a diagonal positive-definite matrix of quadrature weights so that the inner product represents, to a discretization error, the volume-integrated quantity. Note that inner products without any subscripts involve only quadrature weights.  

We define the gain between the forcing and response as the global Rayleigh quotient
\begin{align}
\Psi^2_G(\omega)
= \frac{ \langle \boldsymbol{y},\boldsymbol{y} \rangle_G }{ \langle \boldsymbol{f},\boldsymbol{f} \rangle_G } 
= \frac{ \boldsymbol{f}^H \tc{B}_q^H \tc{R}_G^H \tc{C}_q^H \tc{W}_{G} \tc{C}_q\tc{R}_G\tc{B}_q\boldsymbol{f} }{ \boldsymbol{f}^H \tc{W}_{G} \boldsymbol{f} }. 
\end{align}
The forcing and response that maximize the gain are then sought, and solutions to this standard problem are given by the right- and left-singular vectors and singular values, respectively.  Owing to the weighting matrix, one can either work with a generalized singular value decomposition (SVD), or transform to the standard one by defining $\boldsymbol{k}=\tc{W}_{G}^{1/2}\boldsymbol{f}$ and maximizing
\begin{align}
\Psi^2_G(\omega)
= \frac{ \boldsymbol{k}^H \tc{R}_{GW}^H \tc{R}_{GW} \boldsymbol{k} }{ \boldsymbol{k}^H  \boldsymbol{k} },
\end{align}
where $\tc{R}_{GW} = \tc{W}_{G}^{1/2} \tc{C}_q\tc{R}_G\tc{B}_q \tc{W}_{G}^{-1/2}$ is a weighted form of the resolvent operator.  Optimal gains and forcings are obtained by computing the eigenvalue decomposition of $\tc{R}_{GW}^H \tc{R}_{GW}$ or, equivalently, the SVD of $\tc{R}_{GW}$.  Typically, a direct LU decomposition is performed to factorize the resolvent operator $\tc{R}_G$, and this step constitutes the bulk of the computational cost of resolvent analysis.



\section{Resolvent analysis with OWNS}\label{sec:OptimalOWNS}

In this section, we will use the OWNS-P equations derived in Part 1 to approximate the response of the flow to forcing via spatial marching.  Doing so neglects components of the solution traveling in the upstream direction, which is appropriate only when the base flow is slowly varying in the marching direction; rapid changes in the base flow typically lead to strong upstream reflection/scattering of waves such that neglecting upstream-traveling components of the solution would be unwarranted.  To use the approximate response of OWNS to compute optimal forcing and response pairs, we formulate an adjoint optimization problem whose solution involves upstream marching of adjoint OWNS equations.  


\subsection{The OWNS-P equations}

We begin by briefly reviewing the OWNS-P formulation given in Part 1, with the primary goal of defining necessary variables, before deriving the adjoint optimization procedure in the next section.  

As in the global analysis, the starting point is the compressible linearized Navier-Stokes equations, but unlike~(\ref{eq:LNS1}), the coordinate direction in which the equations are to be parabolized, which we call $x$, remains continuous, and only the transverse directions (normal to $x$) are discretized.  After applying a Fourier transform in time, this leads to the semi-discrete form
\begin{align}
i \omega \boldsymbol{q}  + \tc{A} \dv{\boldsymbol{q}}{x}  + \tc{B}  \boldsymbol{q} & =  \tc{B}_q \boldsymbol{f}.
\label{eq:LNS3}
\end{align}

To aid in identifying upstream- and downstream traveling waves, i.e., those that propagate energy in the negative and positive $x$ directions, respectively, we transform the state vector to characteristic variables
\begin{equation}
\boldsymbol{\phi}(x) = \tc{T}(x)\boldsymbol{q}(x),
\end{equation}
where the rows of $\tc{T}$ are the left eigenvectors of $\tc{A}$ and the diagonal matrix
\begin{equation}
\widetilde{\tc{A}} = \tc{T} \tc{A} \tc{T}^{-1} = \left[  \begin{array}{cc} \tilde{\tc{A}}_{\pm\pm} & \tc{0}  \\ \tc{0} & \tc{0}  \end{array} \right]
\end{equation}
contains the eigenvalues of $\tc{A}$ and $\tilde{\tc{A}}_{\pm\pm}$ contains all of the non-zero eigenvalues. Writing~(\ref{eq:LNS3}) in terms of $\boldsymbol{\phi}$ and solving for $x$-derivatives yields
\begin{subequations}
\label{eq:DAE}
\begin{align}
\tilde{\tc{A}}_{\pm\pm} \frac{d \boldsymbol{{\phi}}_{\pm}}{dx} &= \tc{L}_{\pm\pm} \boldsymbol{{\phi}}_{\pm} + \tc{L}_{\pm0} \boldsymbol{{\phi}}_{0} + \boldsymbol{{f}}_{\phi,\pm}, \label{Eq:DAE_a} \\
\boldsymbol{0} &= \tc{L}_{0\pm} \boldsymbol{{\phi}}_{\pm} + \tc{L}_{00} \boldsymbol{{\phi}}_{0} + \boldsymbol{{f}}_{\phi,0}  \label{Eq:DAE_b},
\end{align}
\end{subequations}
where
\begin{equation}
\label{Eq:defL}
\tc{L} = - \left(i \omega \tc{I} + {\tc{B}} \right) = \left[  \begin{array}{cc} {\tc{L}}_{\pm\pm} & {\tc{L}}_{\pm0}  \\ {\tc{L}}_{0\pm} & {\tc{L}}_{00} \end{array} \right]
\end{equation}
and
\begin{equation}
    \label{eq:f_phi_def}
    \boldsymbol{f_\phi} = \tc{T} \boldsymbol{f}.
\end{equation}
Here, both $\boldsymbol{\phi}$ and $\boldsymbol{f_\phi}$ have been split into parts associated with the nonzero and zero eigenvalues of $\tc{A}$, e.g.,
\begin{equation}
\label{eq:charVars_split}
\boldsymbol{\phi} = \left\{\begin{array}{c} \boldsymbol{\phi}_{\pm} \\ \boldsymbol{\phi}_{0}  \end{array}\right\}.
\end{equation}

The characteristic variable $\boldsymbol{\phi}_{\pm}$ contains contributions from both upstream- and downstream-traveling waves supported by the linearized equations.  The upstream-traveling components can be formally removed using the projection approach derived in Part 1.  The projection operator $\tc{P}$ factorises the solution into
\begin{subequations}
\begin{align}
\textrm{right-going waves:} \quad &\boldsymbol{{\phi}}_{\pm}' = \tc{P} \boldsymbol{{\phi}}_{\pm}, \label{eq:right_going} \\
\textrm{left-going waves:} \quad &\boldsymbol{{\phi}}_{\pm}'' = \left(\tc{I} - \tc{P}\right) \boldsymbol{{\phi}}_{\pm}.
\end{align}
\end{subequations}

Constructing the projection operator $\tc{P}$ requires an eigendecomposition at each $x$, leading to excessive cost.  Instead, a recursive filter was developed in Part 1 that approximates the projection via a set of recursion equations that can be compactly represented via the system
\begin{subequations}
\label{eq:projection_system}
\begin{align}
\boldsymbol{{\phi}}_{\pm}' = \tc{P}_3 \boldsymbol{{\phi}}^{\mathrm{aux}}, \\
\tc{P}_2 \boldsymbol{{\phi}}^{\mathrm{aux}} = \tc{P}_1 \boldsymbol{{\phi}}_{\pm},
\end{align}
\end{subequations}
where $\boldsymbol{{\phi}}^{\mathrm{aux}}$ are a set of auxiliary variables and $\tc{P}_{1,2,3}$ are sparse matrices.  Solving~(\ref{eq:projection_system}) approximates the projection~\ref{eq:right_going} without need for costly eigendecompositions.  

Finally, this approximate projection can be applied to~(\ref{eq:DAE}) to produce a well-posed one-way equation that can be stably integrated in the downstream $x$ direction.  These equations can be compactly expressed as a differential-algebraic input/output system 
\begin{subequations}
\label{Eq:DAE_OWNSP_approx}
\begin{align}
(\tc{A}^\ddagger \partial_x -  \tc{L}^\ddagger)\boldsymbol{\phi}^\ddagger &=   \tc{B}^{\ddagger} \boldsymbol{f_\phi}, \label{Eq:DAE_OWNSP_approx_a} \\
\boldsymbol{{\phi}}' &= \tc{C}^\ddagger \boldsymbol{{\phi}}^\ddagger, \label{Eq:DAE_OWNSP_approx_b}
\end{align}
\end{subequations}
where 
\begin{equation}\label{eq:approx_POWNS}
\boldsymbol{\phi}^\ddagger = \begin{bmatrix} \boldsymbol{{\phi}}_{\pm}' \\ \boldsymbol{{\phi}}_{0}', \\ \boldsymbol{{\phi}}^{\mathrm{aux}} \end{bmatrix}, \quad \tc{A}^\ddagger = \begin{bmatrix} \tc{I} &  &  \\  & \tc{0} &  \\  &  & \tc{0}  \end{bmatrix}, \quad \tc{B}^\ddagger = \begin{bmatrix} \tc{0} & \tc{0} \\ \tc{P}_1\tilde{\tc{A}}_{\pm\pm}^{-1} & \tc{0}  \\ \tc{0} & \tc{I}  \end{bmatrix}\tc{B}_{\boldsymbol{\phi}}, \quad \tc{C}^{\ddagger} = \begin{bmatrix} \tc{I} & \tc{0} & \tc{0} \\ \tc{0} & \tc{I} & \tc{0} \end{bmatrix},
\end{equation}
and
\begin{equation}\label{eq:approx_POWNS_L}
\tc{L}^\ddagger = \begin{bmatrix} \tc{0} & \tc{0} & \tc{P}_3\\ \tc{P}_1\tilde{\tc{A}}_{\pm\pm}^{-1} \tc{L}_{\pm\pm} & \tc{P}_1\tilde{\tc{A}}_{\pm\pm}^{-1} \tc{L}_{\pm0} & -\tc{P}_2 \\ \tc{L}_{0\pm} & \tc{L}_{00} & \tc{0} \end{bmatrix}.
\end{equation}
Notice that here the matrix $\tc{B}^{\ddagger}$ includes the input restriction operator in characteristic variables, $\tc{B}_{\boldsymbol{\phi}} = \tc{T} \tc{B}_q \tc{T^{-1}}$, which was not considered in Part 1. This differential-algebraic equation can be marched efficiently and stably in the $x$-direction with consistent initial condition $\boldsymbol{\phi}_{\pm, in}$ at $x = x_{in}$. The inputs and outputs of the system can be mapped back to the original state and forcing variables as
\begin{equation}
        \boldsymbol{q} = \tc{T}^{-1}\boldsymbol{\phi}', \quad \boldsymbol{f_\phi} = \tc{T} \boldsymbol{f}.
\end{equation}

It should be emphasized that the OWNS approach differs significantly from PSE, which achieve a stable spatial march by numerically damping upstream propagating waves, either by using an implicit axial discretization along with a restriction on the minimum step size~\citep{li1996nature} or by explicitly adding damping terms to the equations~\citep{andersson1998stabilization}. The damping prevents the upstream waves from destabilizing the spatial march, but also introduces uncontrollable errors, to different degrees that depend on the complexity of the solution, into all of the downstream waves \citep{TowneTCFD2018}.

\subsection{Optimal forcing and response with OWNS-P} \label{sec:ResolventOWNS}

The constrained optimization for calculating the optimal response is performed by defining the Lagrangian function
\begin{equation} \label{eq:lag}
    \mathcal{L}(\boldsymbol{\phi}^{\ddagger},{\boldsymbol{\phi}^{\ddagger}}^*,\boldsymbol{\phi}^{\ddagger}_{in},\boldsymbol{\phi}^{\ddagger}_{out},\boldsymbol{f_\phi}) = \mathcal{J}(\boldsymbol{\phi}^{\ddagger}) -\langle {\boldsymbol{\phi}^{\ddagger}}^*,(\tc{A}^\ddagger \partial_x-\tc{L}^\ddagger)\boldsymbol{\phi}^{\ddagger} - \tc{B}^{\ddagger} \boldsymbol{f_\phi} \rangle.
\end{equation}
Here, $\mathcal{J}$ is a generic cost function to be maximized subject to the governing approximate OWNS-P equations, which are enforced via the adjoint variable ${\boldsymbol{\phi}^{\ddagger}}^*$. We define an inner product over the cross-stream coordinates
\begin{align}\label{eq:norm2}
( \boldsymbol{c},\boldsymbol{d} )_O = ( \boldsymbol{c},\tc{W}_e\boldsymbol{d} )	= \boldsymbol{c}^H (x) \tc{W}_{yz} \tc{W}_{e}(x) \boldsymbol{d}(x) &= \boldsymbol{c}^H (x) \tc{W}_{O}(x) \boldsymbol{d}(x), 
\end{align}
and over the entire volume
\begin{align}\label{eq:norm1}
\langle \boldsymbol{a},\boldsymbol{b} \rangle_O = \langle \boldsymbol{a},\tc{W}_e \boldsymbol{b} \rangle	&= \int_\Omega \boldsymbol{a}^H(x) \tc{W}_{yz} \tc{W}_{e}(x) \boldsymbol{b}(x) ~\dd x &= \int_\Omega \boldsymbol{a}^H(x) \tc{W}_{O}(x) \boldsymbol{b}(x) ~\dd x \nonumber \\
& = \int_\Omega ( \boldsymbol{a},\boldsymbol{b} )_O ~\dd x,
\end{align}
where $\Omega = [x_{in},x_{out}]$ is the streamwise extent of the domain, and $x_{in}$ and $x_{out}$ correspond to the streamwise coordinates of the inlet and outlet planes, respectively. As before, the weighting matrix is constructed by composing factors so that the norm represents a quantity of interest (e.g. energy) and quadrature weights for the cross-stream discretization, i.e. $\tc{W}_O(x)=\tc{W}_{yz} \tc{W}_e(x)$.  In this way $\langle \boldsymbol{a},\boldsymbol{b} \rangle_O  \approx \langle \boldsymbol{a},\boldsymbol{b} \rangle_G$ (equal up to a discretization error). 

The inner product in \eqref{eq:lag} can be expanded through integration by parts, yielding 
\begin{align}
   &\langle {\boldsymbol{\phi}^{\ddagger}}^*,(\tc{A^{\ddagger}}\partial_x-\tc{L^{\ddagger}})\boldsymbol{\phi}^{\ddagger} -\tc{B}^{\ddagger} \boldsymbol{f_\phi} \rangle  \nonumber \\
   &= \langle (-\tc{A^{\ddagger}}^*\partial_x-\tc{L^{\ddagger}}^*){\boldsymbol{\phi}^{\ddagger}}^*,\boldsymbol{\phi}^{\ddagger} \rangle + {\left[ \left(\tc{A^{\ddagger}}^*{\boldsymbol{\phi}^{\ddagger}}^*,\boldsymbol{\phi}^{\ddagger}\right) \right]^{x_{out}}_{x_{in}}} -\langle {\boldsymbol{\phi}^{\ddagger}}^*, \tc{B}^{\ddagger} \boldsymbol{f_\phi} \rangle  \nonumber \\ 
   & = \langle (-\tc{A^{\ddagger}}^*\partial_x-\tc{L^{\ddagger}}^*){\boldsymbol{\phi}^{\ddagger}}^*,\boldsymbol{\phi}^{\ddagger} \rangle + \left(\tc{A^{\ddagger}}^*{\boldsymbol{\phi}^{\ddagger}}_{out}^*,{\boldsymbol{\phi}^{\ddagger}}_{out}\right) - \left(\tc{A^{\ddagger}}^*{\boldsymbol{\phi}^{\ddagger}}_{in}^*,{\boldsymbol{\phi}^{\ddagger}}_{in}\right) -\langle {\boldsymbol{\phi}^{\ddagger}}^*, \tc{B}^{\ddagger} \boldsymbol{f_\phi} \rangle, 
       \label{eq:lag3_1}
\end{align}
where
\begin{align}
    & \tc{A^{\ddagger}}^* = {\tc{W}_{yz}^{\ddagger}}^{-1}  \tc{A^{\ddagger}}^H \tc{W}_{yz}^{\ddagger}, \quad \tc{L^{\ddagger}}^* = {\tc{W}_{yz}^{\ddagger}}^{-1}  \tc{L^{\ddagger}}^H \tc{W}^{\ddagger}_{yz}, 
\end{align}
and $\tc{W}_{yz}^{\ddagger}$ is the augmented diagonal quadrature-weight matrix to accommodate the auxiliary variables. Thus, we can define the adjoint to \eqref{eq:approx_POWNS} as
\begin{equation}
    (\tc{A^{\ddagger}}^*\partial_x-\tc{L^{\ddagger}}^*){\boldsymbol{\phi}^{\ddagger}}^*= \boldsymbol{f_\phi}^*.
    \label{eq:adj}
\end{equation}
Note that the negative sign in front of $\tc{A^{\ddagger}}^*$ has been dropped since we are marching upstream and $\boldsymbol{f_\phi}^*$ is the adjoint characteristic forcing determined from the evaluating the stationary points of the Lagrangian function.

The cost function $\mathcal{J}$ is defined similarly to the one of the global resolvent formulation to determine optimal volumetric forcings, $\boldsymbol{f}$, which maximize the energy of the flow in the domain. Here, it is expressed in terms of characteristic variables as%
\begin{align}
\Psi^2_{O}(\omega)
=\mathcal{J}( \boldsymbol{\phi}^{\ddagger},\boldsymbol{f_\phi}) 
=&  \frac{\langle \tc{C}_q \boldsymbol{q},\tc{C}_q \boldsymbol{q} \rangle_O}{\langle \boldsymbol{f},\boldsymbol{f} \rangle_O} = \frac{\int_\Omega  \boldsymbol{q}^H\tc{C}_q^H \tc{W}_{O} \tc{C}_q \boldsymbol{q} ~\dd x}{\int_\Omega (\tc{T}^{-1}\boldsymbol{f_\phi})^H \tc{W}_{O}  (\tc{T}^{-1}\boldsymbol{f_\phi}) ~\dd x} \nonumber \\ 
= &\frac{\int_\Omega {\boldsymbol{\phi}^{\ddagger}}^H \tc{W}^{\ddagger}_{yz} \overbrace{{\tc{W}_{yz}^{\ddagger}}^{-1} \tc{C^{\ddagger}}^H \tc{T}^{-H}\tc{C}_q^H \tc{W}_{O}  \tc{C}_q \tc{T}^{-1}\tc{C^{\ddagger}}}^{\tc{Q}^{\ddagger}_u} \boldsymbol{\phi}^{\ddagger} ~\dd x}{\int_\Omega \boldsymbol{f_\phi}^H \tc{W}_{yz} \underbrace{\tc{W}_{yz}^{-1} \tc{T}^{-H} \tc{W}_{O}  \tc{T}^{-1} }_{\tc{Q}_f} \boldsymbol{f_\phi} ~\dd x} 
= \frac{\langle \boldsymbol{\phi}^{\ddagger},\tc{Q}^\ddagger_u \boldsymbol{\phi}^{\ddagger} \rangle}{\langle \boldsymbol{f_\phi}, \tc{Q}_f \boldsymbol{f_\phi} \rangle}.
\label{eq:JOWNSf}
\end{align}
The Lagrangian function \eqref{eq:lag3_1}, after setting $\boldsymbol{\phi}_{in}=0$, and after substituting the cost function  \eqref{eq:JOWNSf}, is 
\begin{align}
    \mathcal{L}(\boldsymbol{\phi}^{\ddagger},{\boldsymbol{\phi}^{\ddagger}}^*,\boldsymbol{\phi}^{\ddagger}_{out},\boldsymbol{f_\phi}) =  &\frac{\langle \boldsymbol{\phi}^{\ddagger},\tc{Q}^{\ddagger}_u \boldsymbol{\phi}^{\ddagger} \rangle}{\langle \boldsymbol{f_\phi}, \tc{Q}_f \boldsymbol{f_\phi} \rangle} - \langle (\tc{A^{\ddagger}}^*\partial_x-\tc{L^{\ddagger}}^*){\boldsymbol{\phi}^{\ddagger}}^*,\boldsymbol{\phi}^{\ddagger} \rangle 
    \nonumber \\&- \left(\tc{A^{\ddagger}}^*{\boldsymbol{\phi}^{\ddagger}}_{out}^*,\boldsymbol{\phi}^{\ddagger}_{out}\right) + \langle {\boldsymbol{\phi}^{\ddagger}}^*, \tc{B}^{\ddagger} \boldsymbol{f_\phi} \rangle.
    \label{eq:lag3}
\end{align}
The optimal forcing and responses are obtained by finding the stationary points of the Lagrangian function,
\begin{equation}
    \delta \mathcal{L} = \langle \grad_{{\boldsymbol{\phi}^{\ddagger}}^*} \mathcal{L}, \delta {\boldsymbol{\phi}^{\ddagger}}^* \rangle + \langle \grad_{\boldsymbol{\phi}^{\ddagger}} \mathcal{L}, \delta \boldsymbol{\phi}^{\ddagger} \rangle  + \langle \grad_{\boldsymbol{\phi}^{\ddagger}_{out}} \mathcal{L}, \delta \boldsymbol{\phi}^{\ddagger}_{out} \rangle + \langle \grad_{\boldsymbol{f_\phi}} \mathcal{L}, \delta \boldsymbol{f_\phi} \rangle =0.
\end{equation}
We now set each inner product of the stationary points individually to zero,
\begin{subequations}\label{eq:stationary}
\begin{align}
& \langle \grad_{{\boldsymbol{\phi}^{\ddagger}}^*} \mathcal{L}, \delta {{\boldsymbol{\phi}^{\ddagger}}^*} \rangle = 0 \rightarrow \langle -(\tc{A^{\ddagger}}\partial_x-\tc{L^{\ddagger}})\boldsymbol{\phi}^{\ddagger}+{\tc{B}^{\ddagger}}{\boldsymbol{f_\phi}}, \delta {{\boldsymbol{\phi}^{\ddagger}}^*} \rangle = 0, \\
& \langle \grad_{\boldsymbol{\phi}^{\ddagger}} \mathcal{L}, \delta {\boldsymbol{\phi}^{\ddagger}} \rangle = 0 \rightarrow \langle \frac{(\tc{Q}^{\ddagger}_u+{\tc{W}_{yz}^{\ddagger}}^{-1} {\tc{Q}^{\ddagger}_u}^H {\tc{W}_{yz}^{\ddagger}}) {\boldsymbol{\phi}^{\ddagger}}}{\langle \boldsymbol{f_\phi}, \tc{Q}_f \boldsymbol{f_\phi} \rangle}-(\tc{A^{\ddagger}}^*\partial_x-\tc{L^{\ddagger}}^*){{\boldsymbol{\phi}^{\ddagger}}^*}, \delta {\boldsymbol{\phi}^{\ddagger}} \rangle = 0, \\
& \langle \grad_{\boldsymbol{\phi}^{\ddagger}_{out}} \mathcal{L}, \delta {\boldsymbol{\phi}^{\ddagger}_{out}} \rangle = 0 \rightarrow \langle -\tc{A^{\ddagger}}^*{\boldsymbol{\phi}^{\ddagger}}_{out}^*, \delta {\boldsymbol{\phi}^{\ddagger}_{out}} \rangle = 0, \\
& \langle \grad_{\boldsymbol{f_\phi}} \mathcal{L}, \delta \boldsymbol{f_\phi} \rangle = 0, \rightarrow \langle -\frac{\langle {\boldsymbol{\phi}^{\ddagger}},\tc{Q}^{\ddagger}_u {\boldsymbol{\phi}^{\ddagger}} \rangle}{\langle \boldsymbol{f_\phi}, \tc{Q}_f \boldsymbol{f_\phi} \rangle^2}(\tc{Q}_f+\tc{W}_{yz}^{-1} \tc{Q}^H_f \tc{W}_{yz}) \boldsymbol{f_\phi}+ \tc{W}_{yz}^{-1} {\tc{B}^{\ddagger}}^H \tc{W}_{yz}^{\ddagger} {{\boldsymbol{\phi}^{\ddagger}}^*} , \delta \boldsymbol{f_\phi} \rangle = 0. 
\end{align}
\end{subequations}

\begin{figure}
\centering
 \includegraphics[width=0.9\linewidth,trim={0cm 0cm 0cm 0},clip]{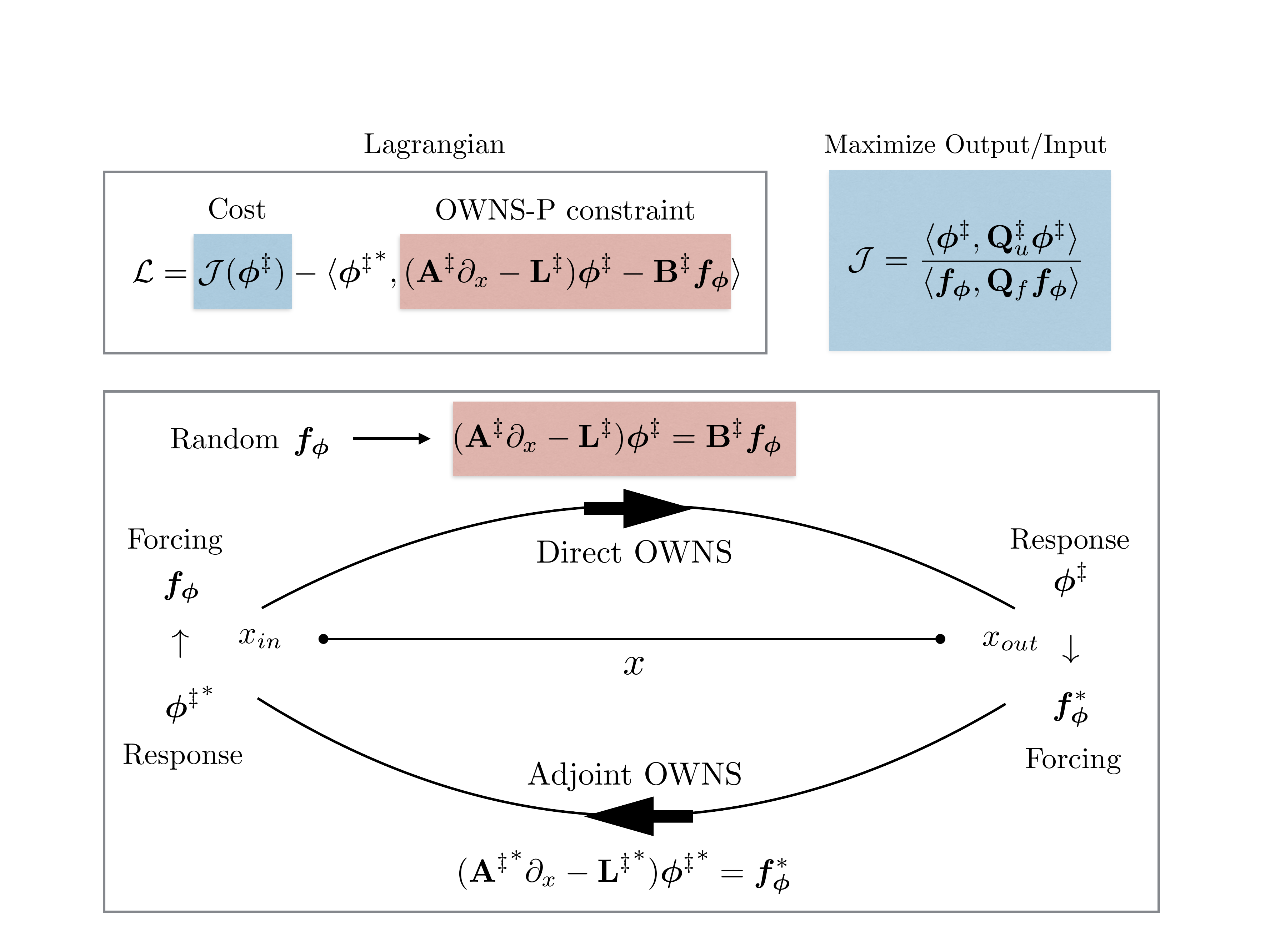}
 \caption{Schematic of the iterative adjoint looping algorithm for the calculation of the optimal responses and forcings using the OWNS framework. By marching the direct and adjoint OWNS-P, instead of the solving the globally discretised equations, a significant computational saving is achieved. }
 \label{fig:OWNScartoon}
\end{figure}

An iterative procedure for finding the stationary points of the system of equations  \eqref{eq:stationary} is summarized in Algorithm \ref{alg:opt3} and depicted schematically in figure~\ref{fig:OWNScartoon}.   Note that $N$ optimal and suboptimal forcings, $\{\boldsymbol{f}_1, \boldsymbol{f}_2, \dots, \boldsymbol{f}_N  \}$, and responses, $\{\boldsymbol{q}_1, \boldsymbol{q}_2, \dots, \boldsymbol{q}_N  \}$, can also be determined by initializing the optimization procedure such that $\langle \boldsymbol{f}_i, \boldsymbol{f}_j \rangle = \delta_{ij}$, and orthonormalizing the forcings at each iteration.

\begin{algorithm}[H]
  \caption{Optimal forcing and response with OWNS}\label{alg:opt3}
  \begin{algorithmic}[1]
\setstretch{1.0}
    \State 
    Initialize random $\boldsymbol{f}_i$ such that $\langle \boldsymbol{f}_i, \boldsymbol{f}_j \rangle_O= \delta_{ij}$. Set threshold $\epsilon$ (here $10^{-4}$).
    
    \Repeat

    \State 
    Solve the forced OWNS-P equations to determine the response by spatial marching in $+x$
    \begin{equation*}
        (\tc{A^{\ddagger}}\partial_x -  \tc{L^{\ddagger}})\boldsymbol{\phi}_i^{\ddagger} = \tc{B}^{\ddagger} {\boldsymbol{f_\phi}}_i,  \quad \boldsymbol{\phi}_i(x=x_{in}) = 0.
    \end{equation*}

    
    \State
    Calculate adjoint forcing 
    $$
    {\boldsymbol{f_\phi}^*}_i = \alpha_i^* (\tc{Q}^{\ddagger}_u+ \tc{W^{\ddagger}}_{yz}^{-1} {\tc{Q}^{\ddagger}_u}^H \tc{W}^{\ddagger}_{yz}) \boldsymbol{\phi}_i^{\ddagger},
    $$
    where $\alpha_i^*$ is a normalization coefficient such that ${\boldsymbol{f_\phi}^*}_i$ has unit norm.
    
    \State 
    Solve the adjoint OWNS-P equations by spatially marching in $-x$
    $$
    (\tc{A^{\ddagger}}^*\partial_x -  \tc{L^{\ddagger}}^*){\boldsymbol{\phi}_i^{\ddagger}}^* = {\boldsymbol{f_\phi}^*}_i,  \quad \boldsymbol{\phi}_i^*(x=x_{out}) =  0.
    $$
    
     \State
    Calculate direct forcing 
    $$
    \boldsymbol{f} =  \mathcal{G} \left[ \tc{T}^{-1}(\tc{Q}_f+ \tc{W}_{yz}^{-1} \tc{Q}^H_f \tc{W}_{yz})^{-1} \tc{W}_{yz}^{-1} {\tc{B}^{\ddagger}}^H {\tc{W}^{\ddagger}}_{yz}{\boldsymbol{\phi}_i^{\ddagger}}^* \right] \label{state:inletBC3}
    $$
    \indent where $\mathcal{G}[ \ ]$ is the Gram-Schmidt orthonormalization operator such that 
    $$
    \langle \boldsymbol{f}_i, \boldsymbol{f}_j \rangle_O = \delta_{ij}.
    $$
    
     \State 
    Calculate the relative change of the cost function
    $$
    \Delta = \left| \frac{\mathcal{J}_{k+1}-\mathcal{J}_{k}}{\mathcal{J}_{k}} \right|=\left|  \frac{\langle \boldsymbol{\q},\boldsymbol{\q} \rangle_{O,{k+1}} - \langle \boldsymbol{\q},\boldsymbol{\q} \rangle_{O,{k}}}{\langle \boldsymbol{\q},\boldsymbol{\q} \rangle_{O,{k}}}\right|.
    $$
    
    \Until{$\Delta<\epsilon$}
    
  \end{algorithmic}
\end{algorithm}


In addition to finding optimal volumetric forcings and responses, optimal initial (boundary) conditions,  as previously done using PSE \citep{tempelmann2010spatial,tempelmann2012spatial}, can be computed by restricting the input to the inlet plane,  prolonging it to the solution space via $\tc{B}_q$ and, if desired, restricting the output to a smaller space, e.g., some downstream plane, via $\tc{C}_q$.  

\section{Resolvent calculations for a jet and a boundary layer}\label{sec:results}

In this section, the optimal OWNS-P algorithm (\S \ref{sec:ResolventOWNS}) is used for the calculation optimal perturbations of a supersonic turbulent round jet at Mach 1.5 and an adiabatic flat-plate boundary layer flow at Mach 4.5.  The method is validated against global resolvent calculations (\S \ref{sec:GlobalResolvent}). All results in this paper are produced using the Caltech Stability and Transition Analysis Toolkit (CSTAT), a comprehensive package capable of performing a wide array of stability calculations from linear stability analysis to planar marching of OWNS and PSE for complex geometries. Details of this code can be found in \cite{kamal_2021}.  For the boundary layer case, the OWNS and global resolvent modes and gain curves at several frequencies and spanwise numbers are also compared against the recent results of \cite{bugeat20193d} for a similar configuration. 
\subsection{Governing equations}

The flow dynamics are governed by the compressible Navier-Stokes equations for an ideal gas
\begin{subequations}
        \begin{align}
          &\frac{D\rho}{D t} +\rho (\nabla \cdot \mathbf{u}) 	= 0, \\
	  &\rho \frac{D \mathbf{u}}{D t} + \nabla p = \frac{1}{Re} \nabla \cdot \tau,  \\
	 &\rho \gamma R \frac{DT}{D t} -(\gamma-1)\frac{Dp}{D t} = \frac{\gamma-1}{Pr Re}\nabla \cdot \left(k\nabla T \right) + \frac{\gamma-1}{Re} \left(\tau:\nabla \mathbf{u}\right),  \\
	 & \tau = 2\mu S  +\kappa (\nabla \cdot \mathbf{u})I, \\
	 & S = \frac{1}{2}\left(\nabla  \mathbf{u} +  (\nabla \mathbf{u})^T - \frac{2}{3} (\nabla \cdot \mathbf{u})I\right), \\
	 & p =  \rho RT, \\
	 & Re = \frac{\rho_\infty a_\infty L}{\mu_\infty}, \\
	 & Pr = \frac{\mu_\infty{c_p}_\infty }{k_\infty}, 
	\end{align}
	\end{subequations}
where $\rho$, $\mathbf{u}$, $t$, $p$, $\gamma$, $R$, $T$, $k$, $\kappa$, $a$, $\mu$, $c_p$, $Re$, and $Pr$ are the density, velocity vector, time, pressure, specific heat ratio, gas constant, temperature, thermal conductivity, bulk viscosity, speed of sound, dynamic viscosity, isobaric specific heat, Reynolds number, and Prandtl number, respectively. All relevant quantities have been non-dimensionalized by the dimensional  ambient quantities $a_\infty$, $\rho_\infty$, $k_\infty$, ${c_p}_\infty$, and $\mu_\infty$, and a problem dependent length-scale $L$. 

In order to obtain the linearized equations for the evolution of the perturbations, the state vector $\q=(\rho, \mathbf{u}, T)^T$
is decomposed into a steady base flow $\qm$ and unsteady fluctuations $\qf$,
\begin{equation}
\q ( \mathbf{x},t) = \qm (\mathbf{x}) + \qf (\mathbf{x},t). 
\end{equation}
The flat plate boundary layer flow is solved in Cartesian coordinates with $\mathbf{x} = (x,y,z)$ corresponding to the streamwise, wall-normal and spanwise directions, respectively and $\mathbf{u}=(u, v, w)$. The jet flow is solved in cylindrical coordinates with   $\mathbf{x} = (x,r,\theta)$ corresponding to the streamwise, radial and azimuthal directions, respectively and $\mathbf{u}=(u_x, u_r, u_\theta)$. Due to the periodicity of the examined flows in the spanwise or azimuthal directions, the perturbation fields are decomposed as $\mathbf{q}'  = \sum \mathbf{\hat{q}}(x,y) \exp (i \beta z + i \omega t)$ and $\mathbf{q}'  = \sum \mathbf{\hat{q}}(x,r) \exp (i m \theta + i \omega t)$, where $\beta$ and $m$ are the spanwise and azimuthal wavenumbers, respectively. 

We assume the fluid to be an ideal gas with $c_v=c_v(T)$ and $c_p = c_p(T)$ and that fluid properties $k$, $\mu$, $\kappa$, and $\gamma$ depend solely on temperature. We denote any of the aforementioned fluid properties as $\Phi$ and perform a Taylor series expansion about $\overline{T}$
\begin{equation}
    \Phi(T) = \Phi(\overline{T}) + \left.\frac{d \Phi}{dT}\right|_{T=\overline{T}}T'+ \text{higher\ order\ terms} .
\end{equation}
The linearized fluid property perturbation is thus
\begin{equation}
\Phi' =  \left.\frac{d \Phi}{dT}\right|_{T=\overline{T}}T'.
\end{equation}
For the present paper, unless otherwise stated, we will assume the fluid as calorically perfect air with $\gamma=1.4$ and $Pr=0.72$ with viscosity and thermal conductivity calculated using Sutherland's law
\begin{equation}
    \mu^\dagger = \mu^\dagger_\infty \left(\frac{T^\dagger}{T^\dagger_\infty}\right)^{3/2} \frac{T^\dagger_\infty + S^\dagger}{T^\dagger + S^\dagger},
    \label{eq:suth}
\end{equation}
where $S^\dagger=110.4$ K and $()^\dagger$ denotes dimensional quantities.

All inner products, e.g., $\langle {\bf a},{\bf b} \rangle_G$, are defined so as to induce the Chu energy norm~\citep{Chu1965}
\begin{equation}
    E_{Chu} = \frac{1}{2}\int \int_\Omega \frac{R \overline{T}}{ \overline{\rho}} \rho'^2 + \overline{\rho} |\mathbf{u'}|^2 + \frac{R \overline{\rho}}{\overline{T}(\gamma(\overline{T})-1) }T'^2 ~\dd x \dd A \approx \int_\Omega \vecd{q'}^*\tc{W}_{yz}  \tc{W}_e  \vecd{q'}~\dd x,
\end{equation}
where $\tc{W}_e$ is the diagonal Chu energy weight matrix defined as
\begin{equation}
 \tc{W}_e = \frac{1}{2}\begin{bmatrix} \frac{R \overline{T}}{ \overline{\rho}} & 0 & 0 & 0	& 0  	\\  0  & \overline{\rho} & 0 & 0 & 0 \\ 0 & 0 &  \overline{\rho} & 0 & 0 \\ 0 & 0 & 0 &  \overline{\rho} & 0 \\ 0 & 0 & 0 & 0 & \frac{R \overline{\rho}}{\overline{T}(\gamma(\overline{T})-1) }\end{bmatrix}.
\end{equation}

\subsection{Base flow calculation}

\begin{figure}
\includegraphics[width=1\linewidth,trim={6cm 3.2cm 6cm 0},clip]{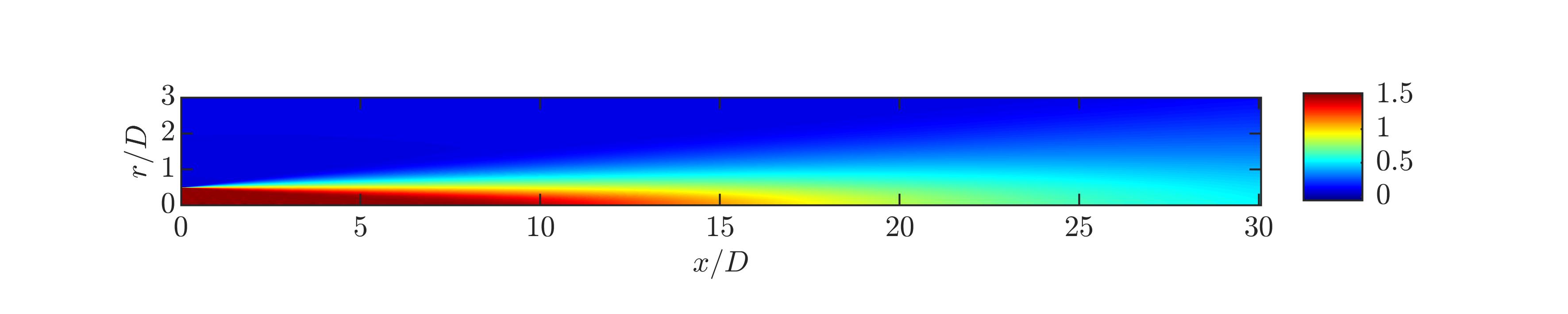}
 \caption{Mean flow of the M=1.5 turbulent jet flow. Local Mach number is shown.}
 \label{fig:Jetbaseflow}
\end{figure}

 The mean flow for a $M = U_{j} / c_{\infty} = 1.5$ turbulent axisymmetric jet is obtained from a large-eddy simulation \citep{Bres2017unstructured}, described also in Part 1. Here, the subscript $j$ denotes conditions at the jet nozzle exit. The mean flow is shown in  figure \ref{fig:Jetbaseflow}.

\begin{figure}
\includegraphics[width=0.47\linewidth,trim={0cm 0cm 0cm 0},clip]{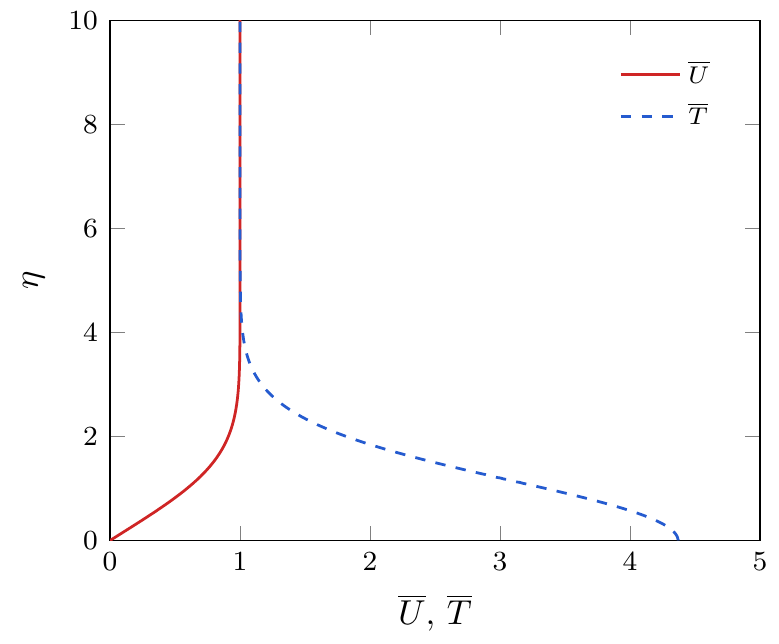} 
\includegraphics[width=0.57\linewidth,trim={5.cm 0cm 6cm 0},clip]{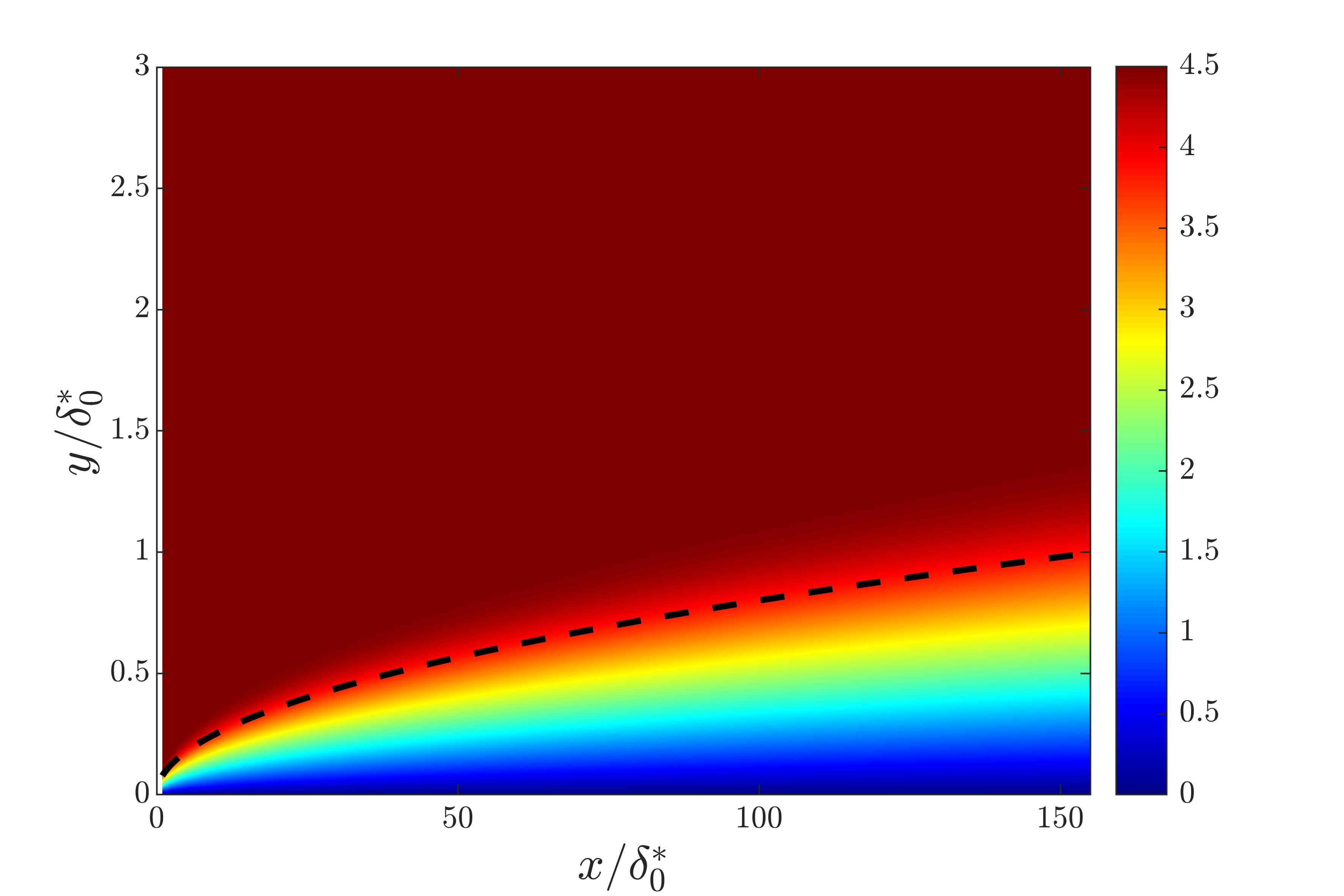}
 \caption{Base flow of the $M=4.5$ adiabatic flat-plate boundary layer obtained from a self-similar transformation.  Self-similar streamwise and temperature components (left). Local Mach number (right) with the dashed line corresponding to the displacement thickness.}
 \label{fig:BLbaseflow}
\end{figure}

 The laminar base flow for a $M=U_{\infty} / c_{\infty}=4.5$ flat-plate zero-pressure gradient boundary layer is obtained from a  self-similar solution of the compressible boundary layer equations. The Howarth-Dorodnitsyn transformation \citep{Stewartson1964} was employed to reduce the governing equations to a set of ODE's, given in appendix \ref{app:baseflow}.  The self-similar solution is shown in  figure \ref{fig:BLbaseflow}.
All the boundary layer quantities are non-dimensionalized using the freestream velocity $U_{\infty}$ and the local compressible displacement thickness $\delta^*(x)$, or the displacement thickness at the outlet of the domain $\delta^*_0$. The outlet of the domain is at  $Re_x^{out} = \frac{\nu_\infty x}{U_{\infty}} = 1.74\times 10^6$, or $Re_{\delta^*_0}^{out}=11,216$.  The domain inlet is located a small distance form the leading edge at $Re_x^{in} = 105$, corresponding to $Re_{\delta^*_0}^{in}=871$. The domain size is similar to the one of \cite{bugeat20193d}, where in their calculations the outlet is at $Re_x^{out} = 1.75\times 10^6$, or $Re_{\delta^*_0}^{out}=11,000$. However, in their calculations the flat plate leading edge was also included in the computational domain resulting in a weak shock at the leading edge and small discrepancies in the momentum thickness at the outlet compared to the self-similar solution that we use in this study. 

\subsection{OWNS \& global solver setup}
The OWNS equations are discretized in the transverse direction using fourth-order central finite differences with summation-by-parts boundary closure \citep{Strand1994summation}. Integration of the OWNS equations in $x$ is performed using a backward differentiation formula of order 2 (BDF2). Far-field radiation boundary conditions are enforced at the free transverse boundary by Thompson characteristic conditions \citep{thompson1987time}.  

The global equations are discretized in the transverse and streamwise directions using fourth-order central finite differences with summation-by-parts boundary closure. Streamwise boundary conditions are implemented which include inlet and outlet sponges in addition to local inviscid characteristic boundary conditions~\citep{thompson1987time}.

For the boundary layer, in order to properly resolve the instabilities near the wall and critical layer, grid stretching in the $y$-direction is employed clustering half of the points near the wall for $y/{\delta^*_0}<0.9$ \citep{Malik1990}. Isothermal no-slip boundary conditions ($\hat{u}=\hat{v}=\hat{w}=\hat{T}=0$) are enforced on the wall~\citep{poinsot1992boundary}.

\begin{figure}
 \includegraphics[width=1\linewidth,trim={0cm 0cm 0cm 0},clip]{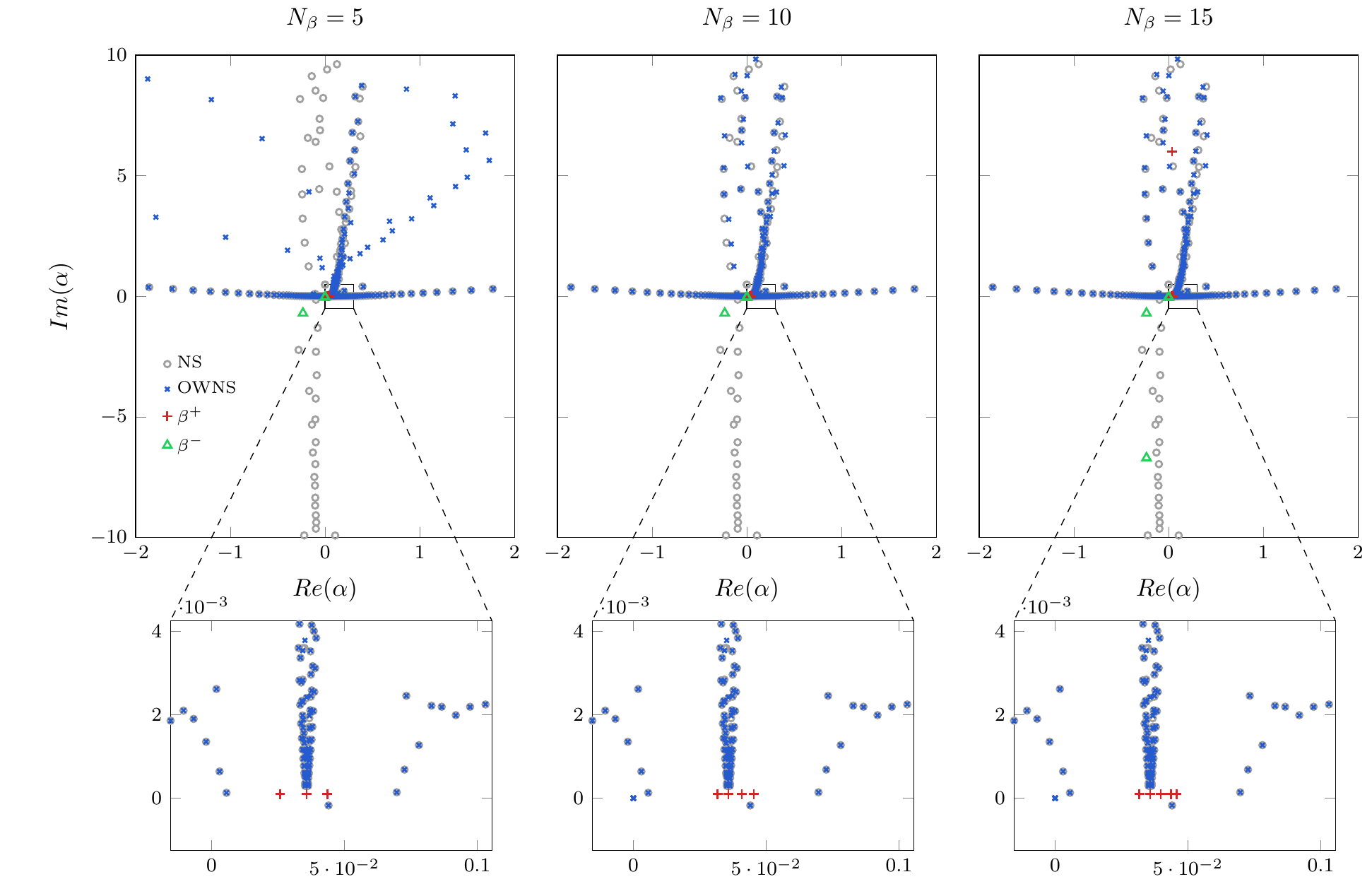}
 \caption{Placement of recursion parameters $\beta^+$ and $\beta^-$ for the $M=4.5$ flat-plate boundary layer flow. The local  spectra of the Navier-Stokes and one-way Navier-Stokes operators are shown. The $\beta^+$ recursion parameters are placed near the downstream propagating modes, whereas the $\beta^-$ near the upstream acoustic ones.  For a sufficiently large number $N_\beta$ of recursion parameters, the OWNS operator converges to the downstream propagating modes, thus eliminating the ellipticity associated with upstream modes and enabling stable, convergent marching in the slowly-varying $x$ direction. }
 \label{fig:recursions}
\end{figure}


An important part in the calculation of the OWNS response is the placement of the recursion parameters. An estimate for the eigenvalues of the downstream and upstream modes can be obtained assuming locally parallel and uniform flow. A recipe  assuming inviscid flow (Euler equations) has been presented in  \cite{towne2015one}. Here, a similar procedure is followed for the choice of the recursion parameters for wall bounded flows (Navier Stokes equations). The estimate at each streamwise location is based on the freestream velocity and frequency of the perturbation. All calculations were performed with $N_{\beta}=15$, which gave a good approximation for the filtering of the upstream propagating modes and accurate capturing of the downstream ones. The effectiveness of the OWNS-P to accurately filter the upstream propagating modes without modifying the downstream ones is demonstrated in figure  \ref{fig:recursions} where the local (in $x$) spectra of the Navier-Stokes and OWNS-P equations are shown. The spectra were calculated by solving a generalised eigenvalue problem for the unforced (homogeneous) equations given in \eqref{eq:DAE} and \eqref{eq:approx_POWNS} assuming a locally parallel flow, where $\partial_x \to i \alpha$ and $\alpha$ corresponds to the complex eigenvalue in the streamwise direction (wavenumber and growth rate).

\setlength{\tabcolsep}{8pt}
\renewcommand{\arraystretch}{1.5}
\begin{table}
  \begin{center}
\def~{\hphantom{0}}
\begin{tabular}{l c c c c c c c c}
Jet & & $St$ & $m$ & $N_x \times N_r$ & $N_x \times N_r$ & $x/D$ &  $r/D$\\ 
      & &          &         & (OWNS)         & (Global) \\ 

 \hline
  Kelvin Helmholtz &   & 0.26 & 0 & $591 \times 250$ &  $701 \times 250$ & $0.5-30$ & $0-17$\\
  Kelvin Helmholtz &   & 0.52 & 0 & $591 \times 250$ &  $701 \times 250$ & $0.5-30$ & $0-17$\\
 \hline \hline
  Boundary layer & & $\omega$ & $\beta$ & $N_x \times N_y$ & $N_x \times N_y$ & $x/\delta^*_0$ &  $y/\delta^*_0$\\ 
        & &          &         & (OWNS)         & (Global) \\ 
 \hline
  Streaks &   & 0.002 & 2.2 & $1601 \times 81$ & $601 \times 108$ & $0.01-155$ & $0-9$\\
  Second mode &   & 2.5 & 0 & $2401 \times 108$ & $1001 \times 180$ & $0.01-155$ & $0-9$\\
  Oblique mode &   & 0.32 & 1.2 & $1601 \times 81$ & $651 \times 81$ & $0.01-155$ & $0-9$\\
\end{tabular}
\caption{Numerical parameters for the jet and boundary layer cases.}
  \label{tab:param}
  \end{center}
\end{table}

Optimal forcing and response modes have been computed using the OWNS-P and global methods for several different frequency and wavenumber combinations corresponding to different instability mechanisms for the jet and boundary layer flows.  These cases, and the numerical parameters uses for each, are summarized in table \ref{tab:param}.


\subsection{Jet optimal perturbations}\label{sec:}

\begin{figure}
\begin{center}
\hspace{0cm} {$f_p: Global$, $St=0.26$} \hspace{3cm} {$p': Global$} \hspace{0 cm}  \\
\includegraphics[width=0.48\linewidth,trim={9cm 10.8cm 12cm 8.8cm},clip]{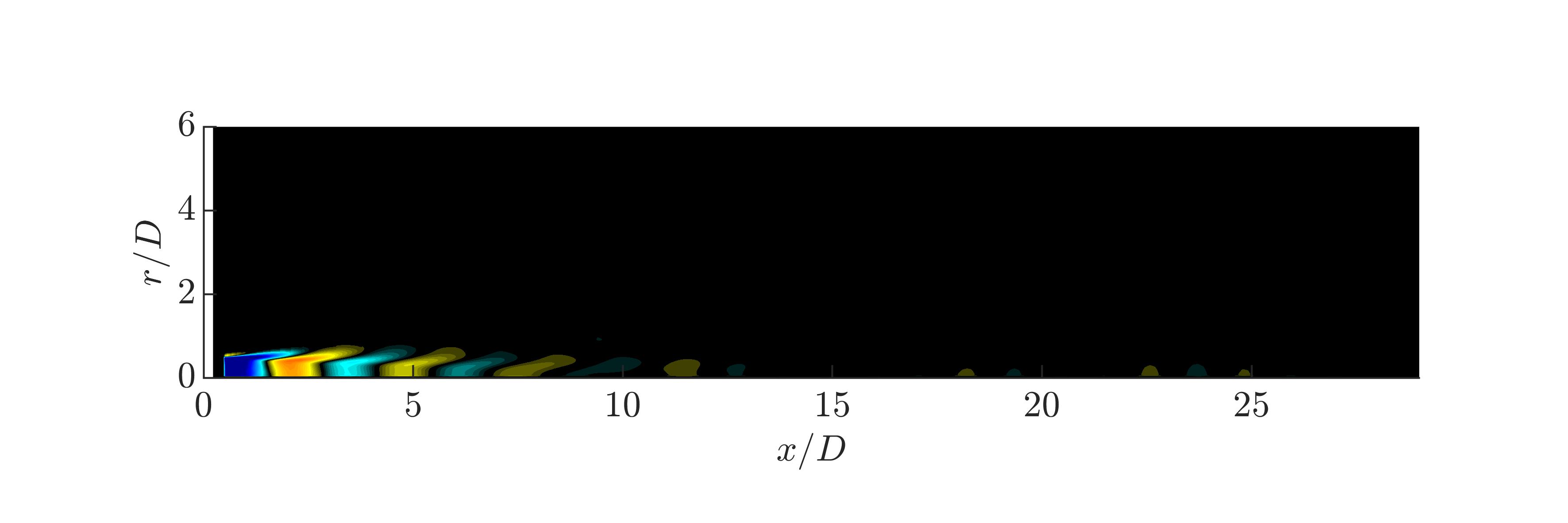}
\includegraphics[width=0.48\linewidth,trim={9cm 10.8cm 12cm 8.8cm},clip]{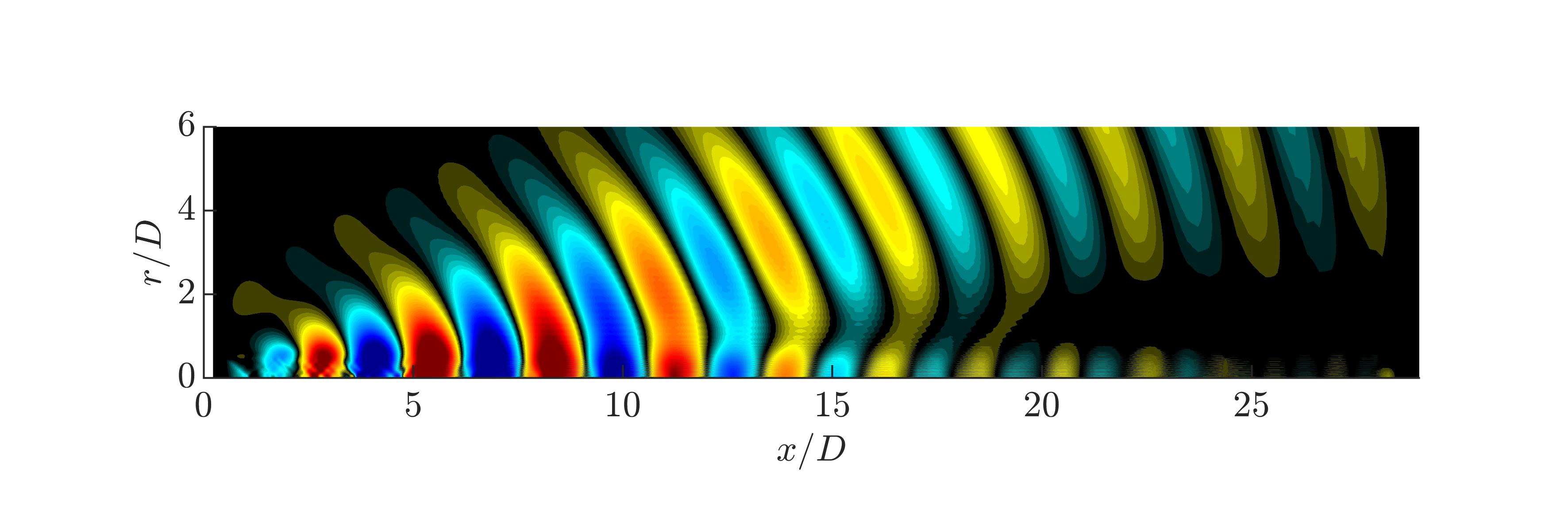}\\
\hspace{0cm} {$f_p: OWNS$, $St=0.26$} \hspace{3cm} {$p': OWNS$} \hspace{0 cm}  \\
\includegraphics[width=0.48\linewidth,trim={9cm 10.8cm 12cm 8.8cm},clip]{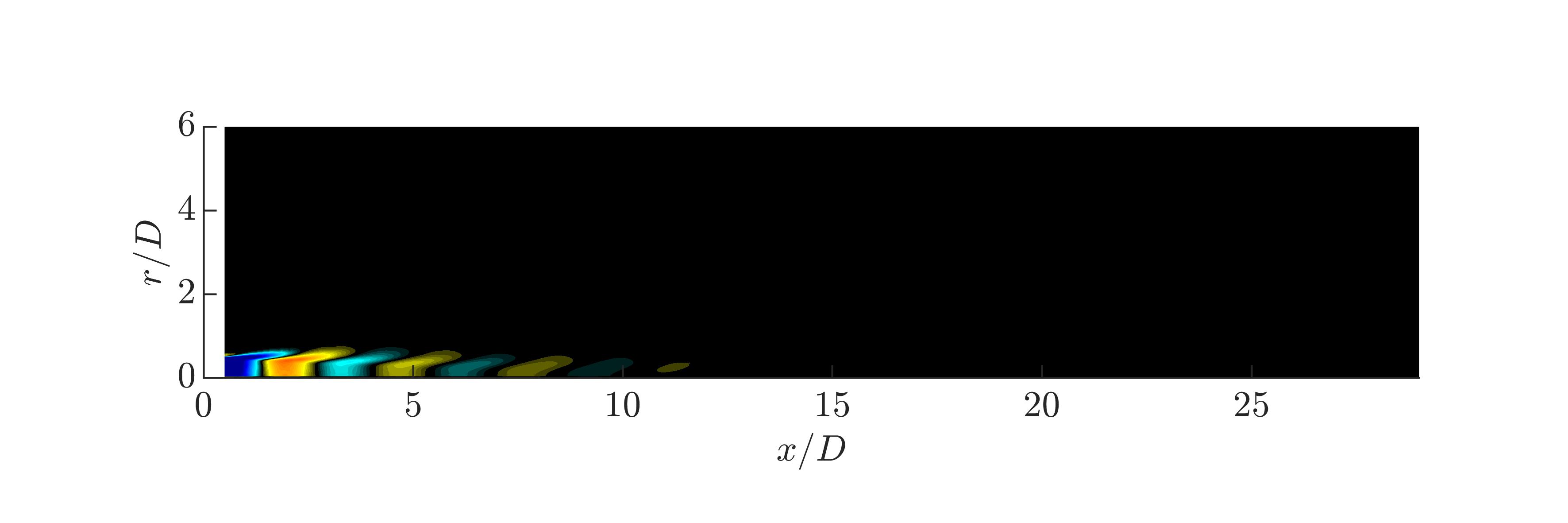}
\includegraphics[width=0.48\linewidth,trim={9cm 10.8cm 12cm 8.8cm},clip]{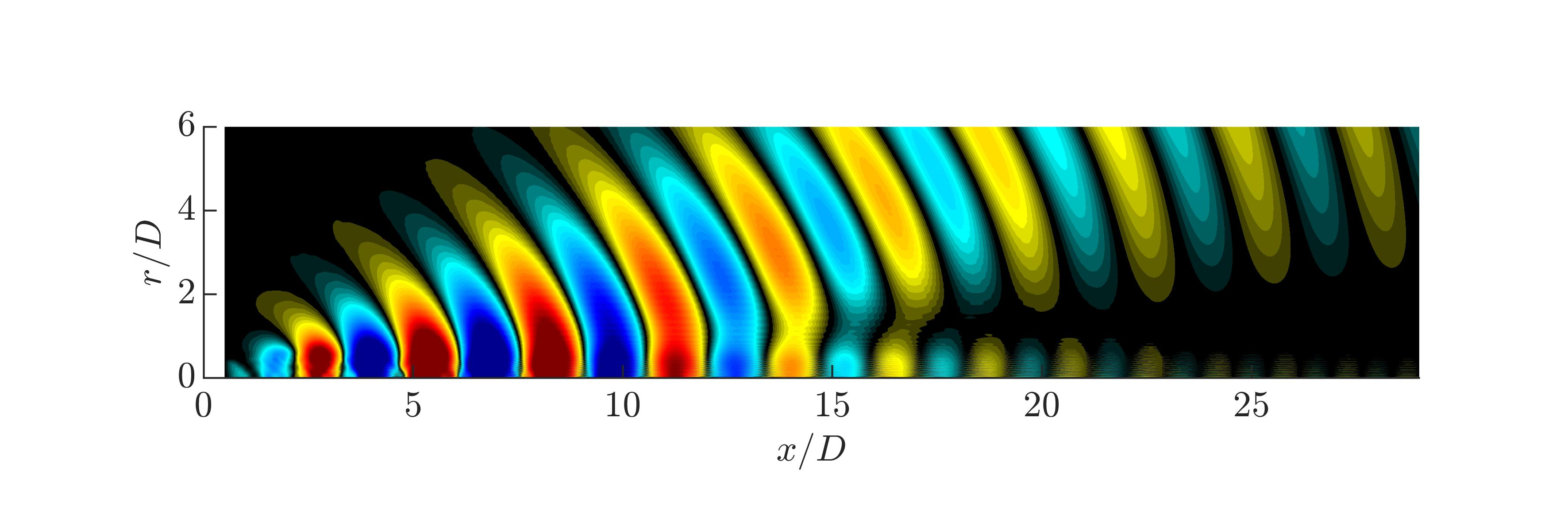}\\
\hspace{0cm} {$f_p: Global$, $St=0.52$} \hspace{3cm} {$p': Global$} \hspace{0 cm}  \\
\includegraphics[width=0.48\linewidth,trim={9cm 10.8cm 12cm 8.8cm},clip]{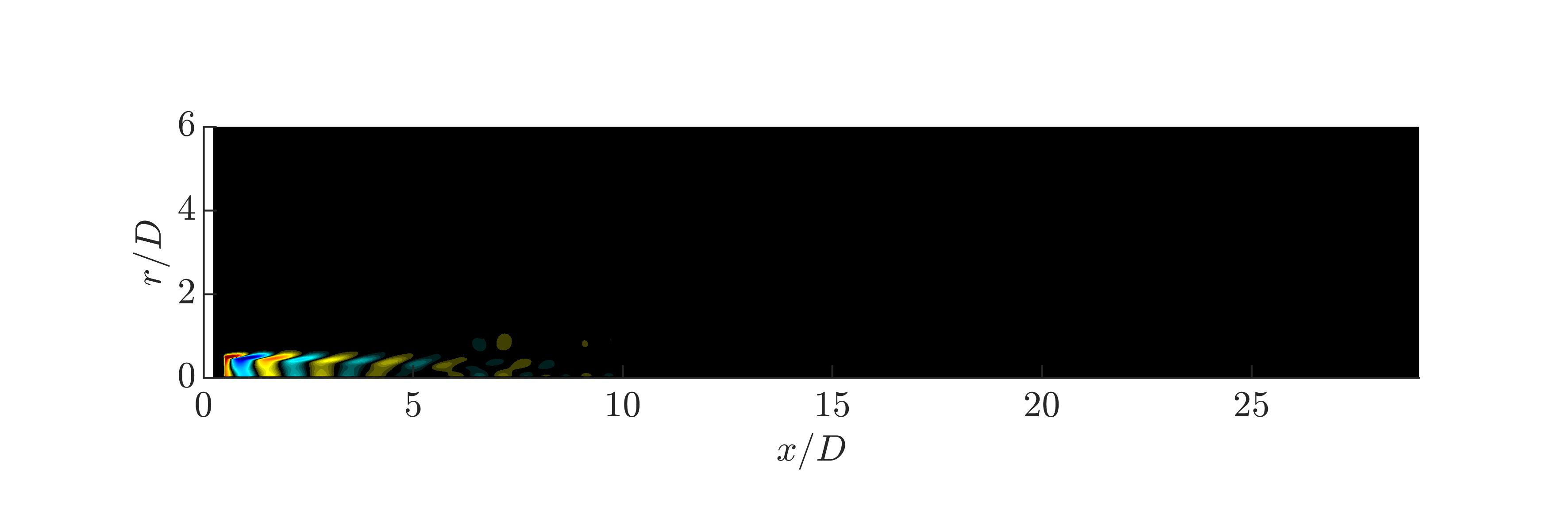}
\includegraphics[width=0.48\linewidth,trim={9cm 10.8cm 12cm 8.8cm},clip]{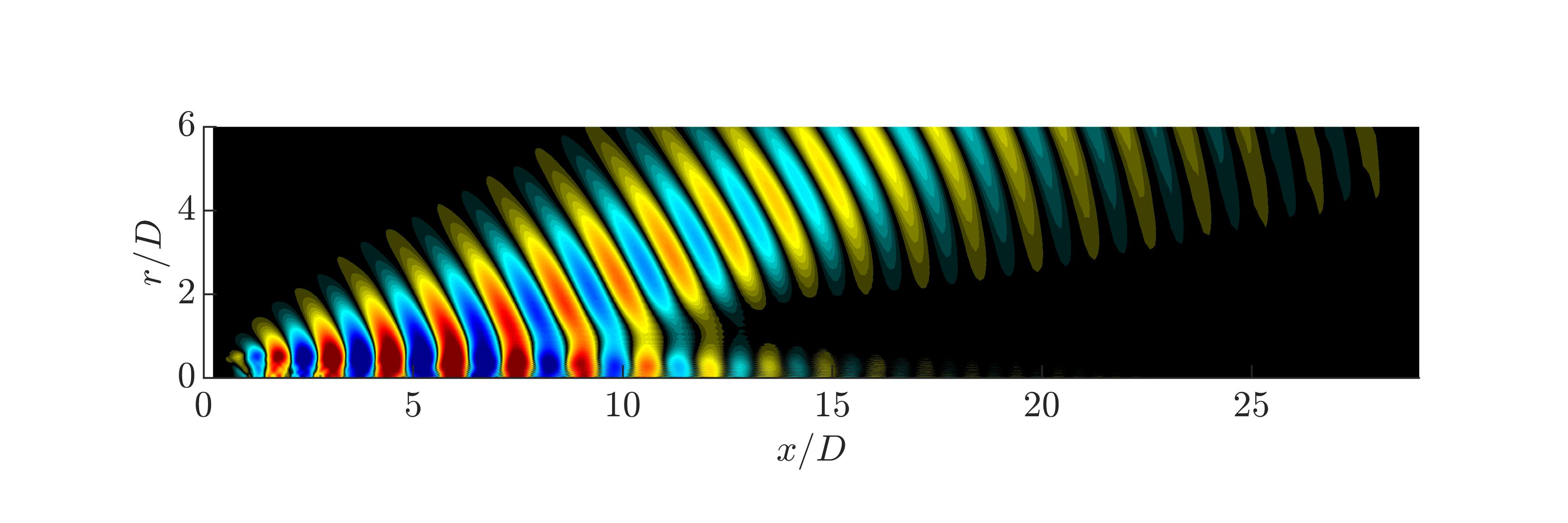}\\
\hspace{0cm} {$f_p: OWNS$, $St=0.52$} \hspace{3cm} {$p': OWNS$} \hspace{0 cm}  \\
\includegraphics[width=0.48\linewidth,trim={9cm 5cm 12cm 8.8cm},clip]{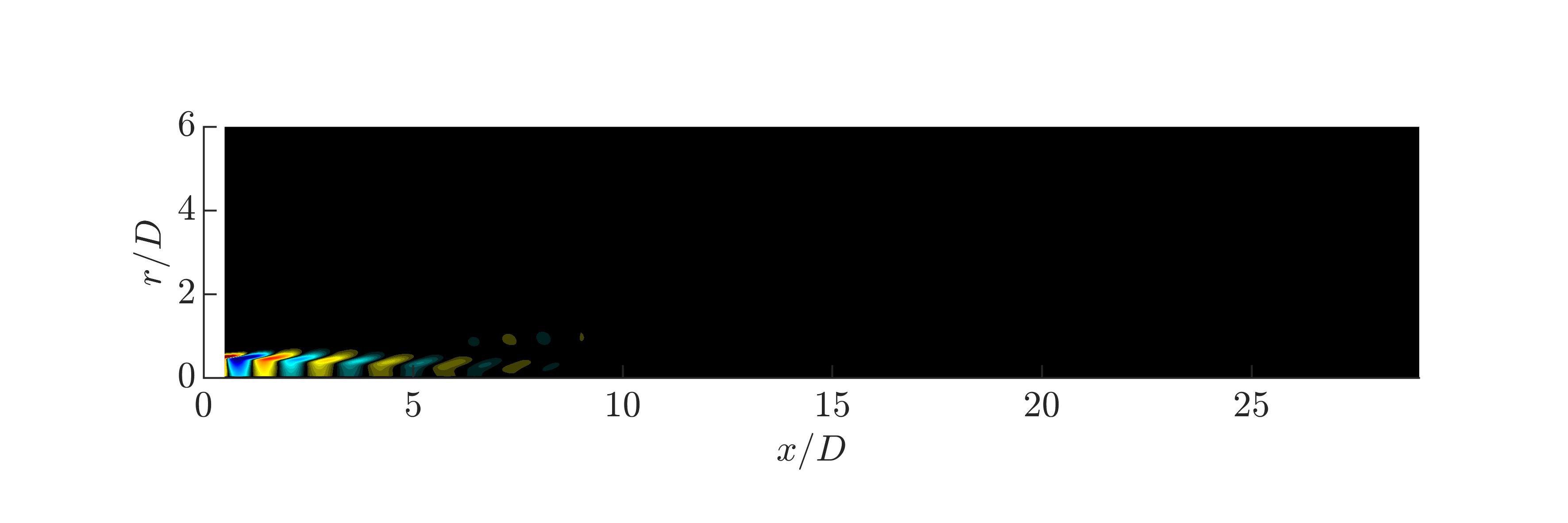}
\includegraphics[width=0.48\linewidth,trim={9cm 5cm 12cm 8.8cm},clip]{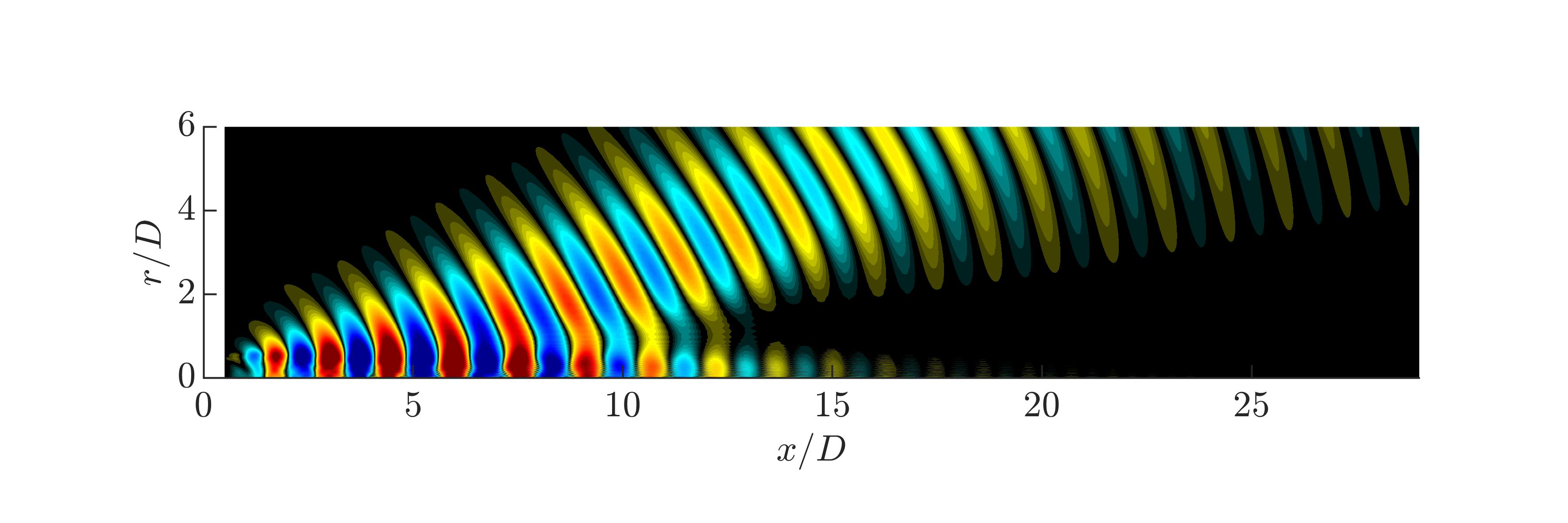}
\end{center}
 \caption{Optimal forcing and response for the axisymmetric turbulent $M=1.5$ jet. Comparison between global and OWNS. Contours of the real pressure component at $St=0.26$ and $St=0.52$, both with $m=0$, are shown.}
 \label{fig:jetcontour}
\end{figure}

First, we compute the optimal forcing and response of the turbulent jet flow using the global resolvent framework (\S \ref{sec:GlobalResolvent}) for two different frequencies, corresponding to $St=0.26$ and $St=0.52$ with  azimuthal wavenumber $m=0$. The linearized equations have been globally  discretised in the inhomogeneous directions $x$ and $r$, as described in the previous section. This corresponds to a similar procedure as that followed by \cite{SchmidtJFM2018}. At these frequencies, the response is dominated by the amplification of disturbances due to the Kelvin-Helmholtz instability \citep{pickering2020lift}. 

The input forcing has been constrained to $0.5 < x/D < 30$, $R_{min} < r/D < R_{max}$. We use $R_{min}=0.0425$ to prevent the forcing from damaging the pole conditions \cite{Mohseni2002evaluation} and $R_{max}$ is defined as the radial jet location where the velocity is greater than $5\%$ of the maximum jet velocity, i.e., the forcing is contained only within the jet.  The output is defined in the full numerical domain, $0.5 < x/D < 30, 0 < r/D < 17$.


The optimal forcing and response for the same two frequencies are also calculated using the optimal OWNS-P methodology (\S \ref{sec:ResolventOWNS}). The linearized equations are discretized in the inhomogeneous direction $r$, whereas in the slowly varying $x$ direction, spatial marching is employed for the calculation of the direct and adjoint solutions reducing substantially the computational cost when compared to the global method.


The real part of the optimal pressure forcing and response from the OWNS-P and global resolvent methods are shown in figure \ref{fig:jetcontour}. Quantitative agreement is obtained, justifying the assumption of weak influence of the upstream propagating modes during the OWNS-P parabolization procedure.

\subsection{Boundary layer optimal perturbations}

\begin{figure}
\centering
\includegraphics[width=1\linewidth,trim={0cm 0cm 00cm 0},clip]{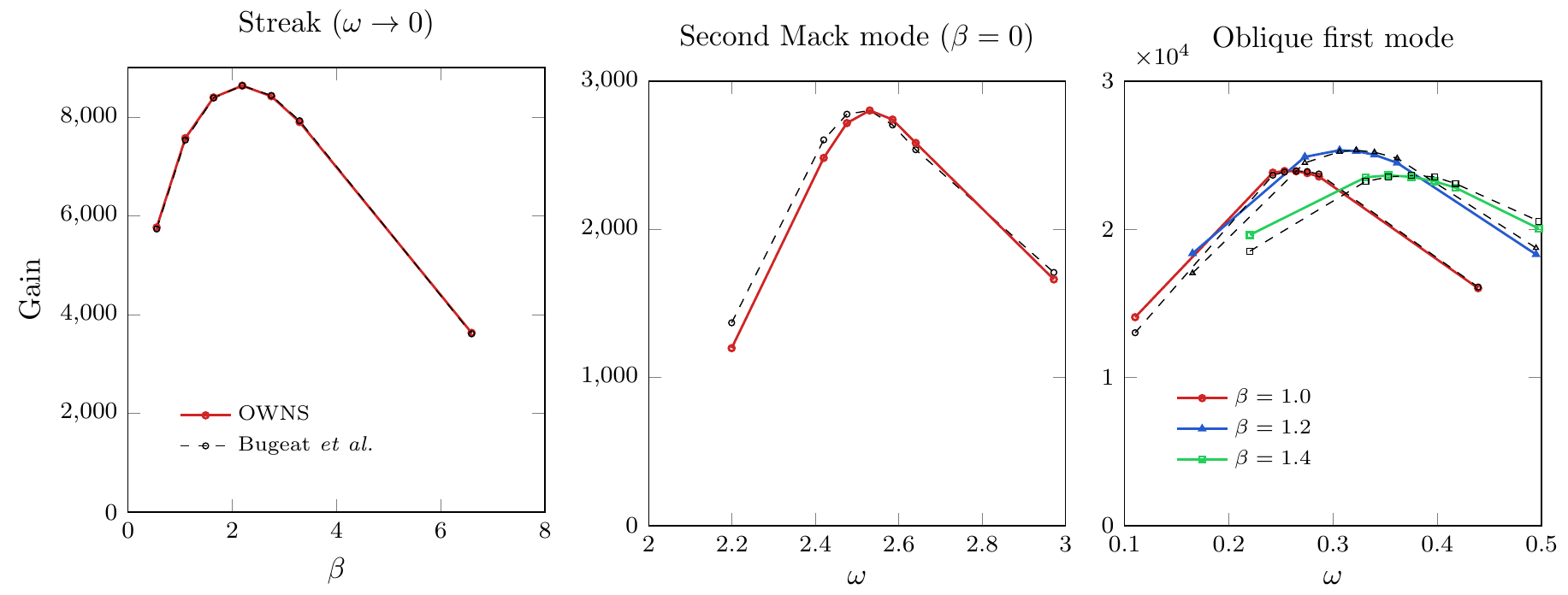}
 \caption{Optimal input/output gain for three linear instability mechanisms for the M=4.5 flat plate adiabatic boundary layer. They are identified as regions of locally maximum gain in the $\beta-\omega$ plane. Validation against normalized results from \cite{bugeat20193d} for a similar configuration.}
 \label{fig:Gains}
\end{figure}

The resolvent modes for the adiabatic flat plate boundary layer have been calculated with the optimal OWNS-P method for a range of frequencies and spanwise wavenumbers. In figure \ref{fig:Gains}, the optimal input/output gains corresponding to three regions of locally maximum gain in the $\beta-\omega$ plane are shown. Their maxima correspond to the amplification of streaks ($\omega \to 0$), second Mack modes ($\beta =0$) and oblique first modes. For all three instability types, we observe excellent agreement of the optimal $\omega$ and/or $\beta$ when compared to the global resolvent calculations of \cite{bugeat20193d}. For all the calculations presented here, the forcing has been restricted only to the momentum components $(f_u,f_v,f_w)$ to match the setup of \cite{bugeat20193d}.

\begin{figure}
\begin{center}
\hspace{0cm} {$f_v: Global$} \hspace{4cm} {$u': Global$} \hspace{0 cm}  \\
\includegraphics[width=0.48\linewidth,trim={9cm 7cm 22cm 2cm},clip]{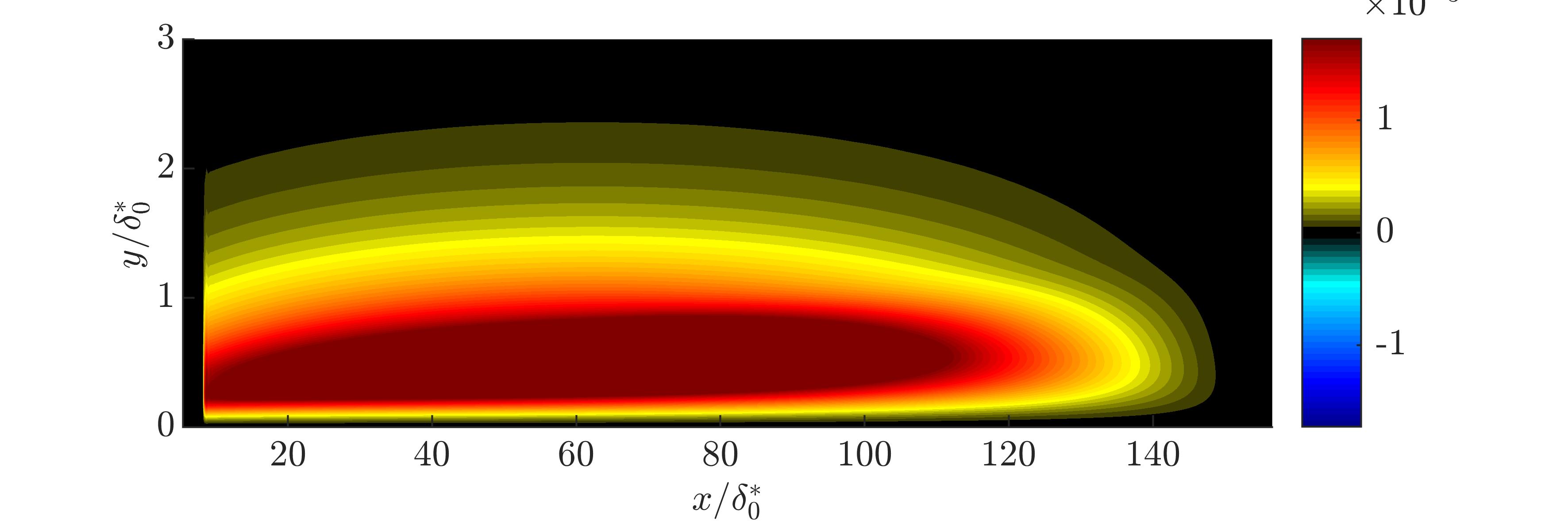}
\includegraphics[width=0.48\linewidth,trim={9cm 7cm 22cm 2cm},clip]{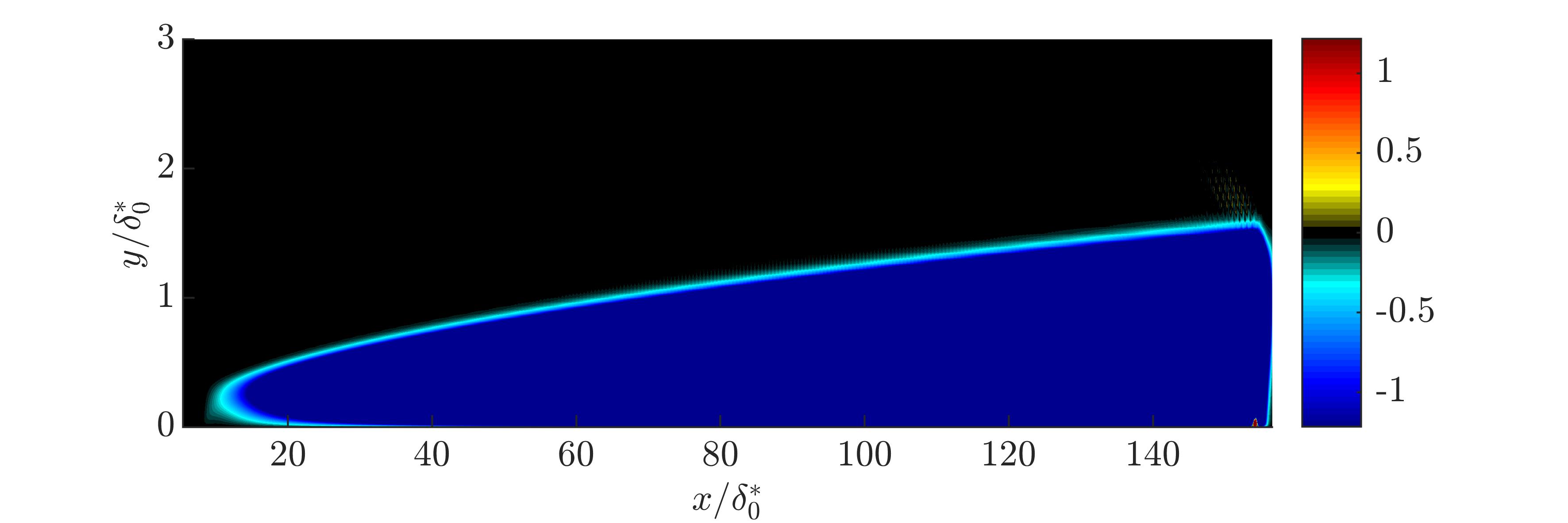}
\\
\hspace{0cm} {$f_v: OWNS$} \hspace{4cm} {$u': OWNS$} \hspace{0 cm}
\\
\includegraphics[width=0.48\linewidth,trim={9cm 0cm 22cm 2cm},clip]{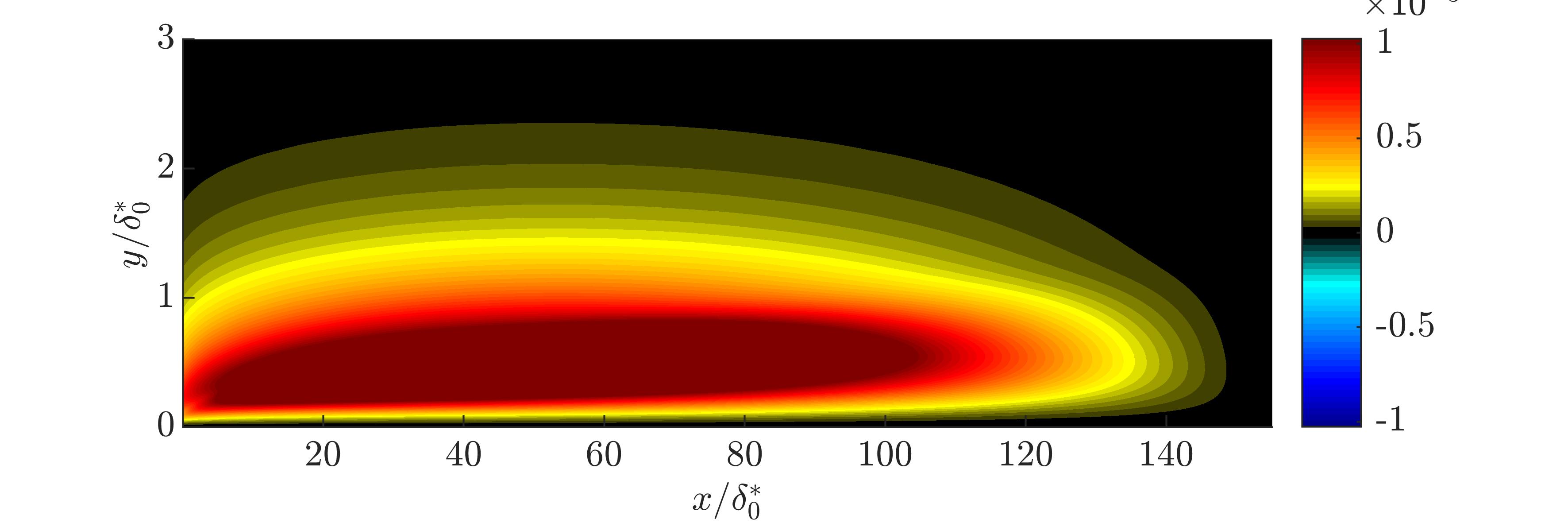}
\includegraphics[width=0.48\linewidth,trim={9cm 0cm 22cm 2cm},clip]{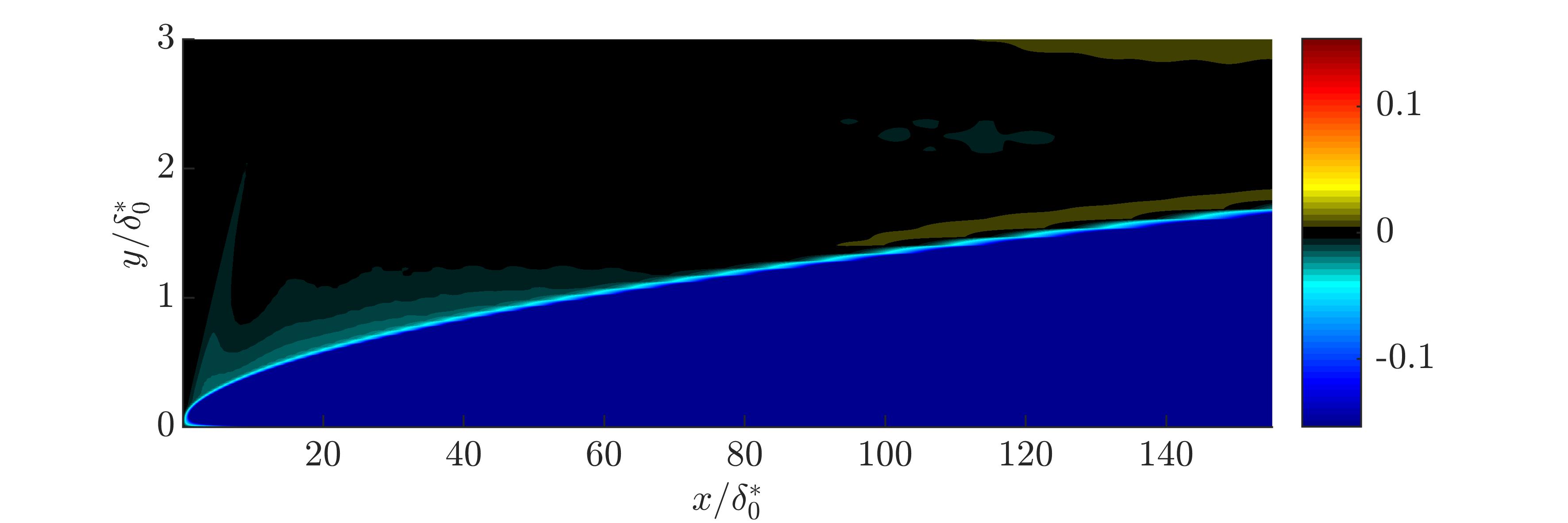}\\
\vspace{1cm}
 \includegraphics[width=0.4\linewidth,trim={0cm 0cm 0cm 0},clip]{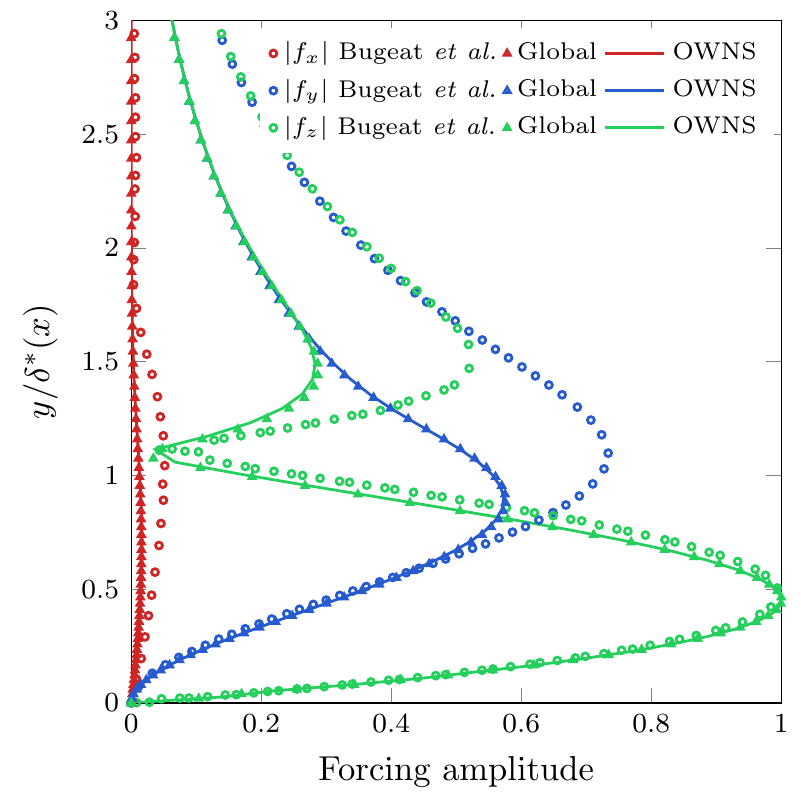}
 \hspace{1cm}
  \includegraphics[width=0.4\linewidth,trim={0cm 0cm 0cm 0},clip]{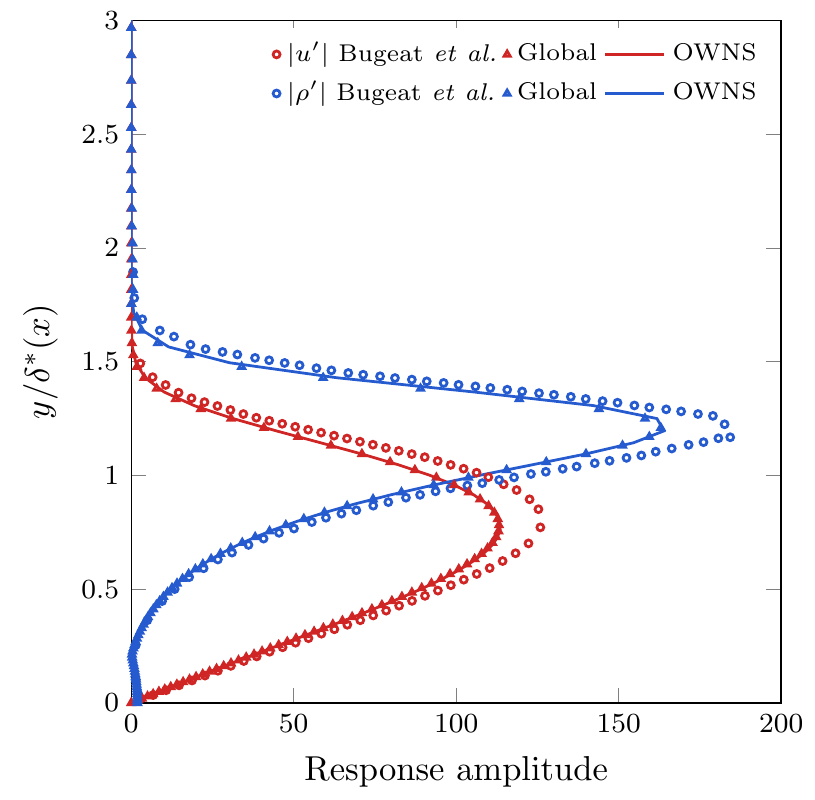}
\end{center}
 \caption{Streak (steady 3D) optimal disturbances at $\omega=0.002$, $\beta=2.2$. Forcing and response amplitude components at $x/\delta^*_0=35$ (left) and $x/\delta^*_0=159$ (right). Circle symbols: \cite{bugeat20193d}; triangle symbols: global; solid lines: OWNS.}
 \label{fig:streaks}
\end{figure}

For streaks, the optimal forcing consists of streamwise counter-rotating vortices that lift the streamwise base flow momentum. This is referred to as the \emph{lift-up} mechanism and yields a response that contains primarily streaks of highly amplified streamwise velocity stretching in the streamwise direction. The dominant input ($f_v$) and output ($\hat{u}$) velocity modes from OWNS-P and global methods are shown in figure \ref{fig:streaks}. Quantitative agreement is again obtained between the OWNS-P optimal modes and the global ones, but minor differences are observed near the inlet/outlet boundaries of the domain, where the sponges of the global method attenuate the response to avoid reflections. Based on our experience, the tuning of the sponges is a cumbersome procedure and problem specific, a step that is bypassed during the OWNS-P marching since upstream propagating wave have been eliminated at each $x$ location during the parabolization procedure. 

In figure \ref{fig:streaks}, we also plot the OWNS-P optimal forcing and response profiles for all the perturbation components at a fixed streamwise location near the  inlet and outlet, respectively. The OWNS-P results are compared against our global calculations for exactly the same configuration and the global calculations from \cite{bugeat20193d}. Quantitative agreement is again obtained between OWNS-P and our global results. Minor discrepancies are observed for the forcing far from the wall when our results are compared against \cite{bugeat20193d}. These discrepancies can be attributed to the differences in the computational domain (inclusion of leading edge and shock) and the different choice of the input norm between the two studies. Specifically for the latter, \cite{bugeat20193d} define the energy norm of the input (forcing) based solely on the spatial quadrature weights, whereas here we have included the Chu definition for the energy. 

\begin{figure}
\begin{center}
\hspace{0cm} {$f_w: Global$} \hspace{4cm} {$\hat{u}: Global$} \hspace{0 cm}  \\
\includegraphics[width=0.48\linewidth,trim={9cm 7cm 22cm 2cm},clip]{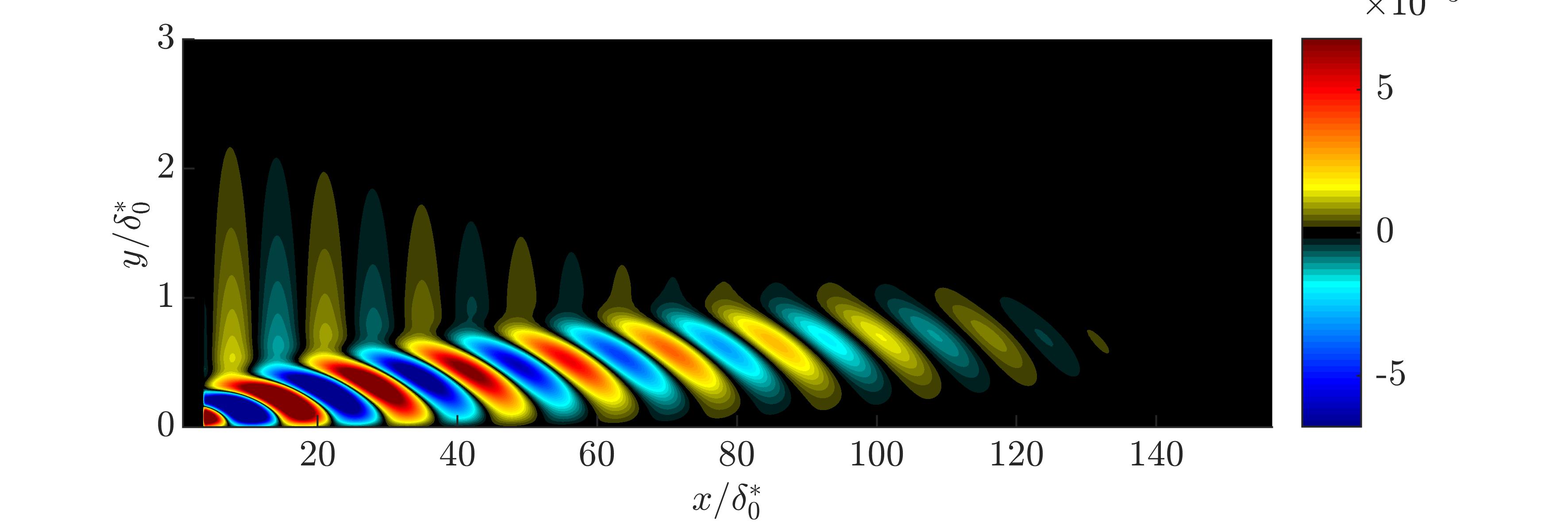}
\includegraphics[width=0.48\linewidth,trim={9cm 7cm 22cm 2cm},clip]{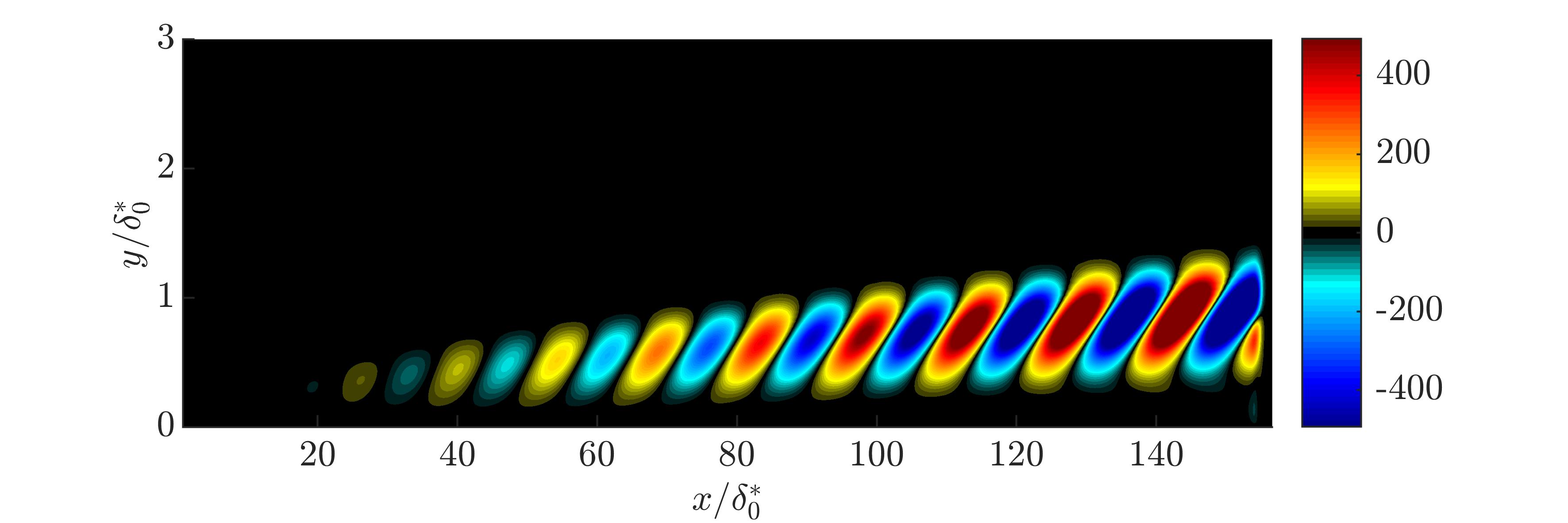}
\\
\hspace{0cm} {$f_w: OWNS$} \hspace{4cm} {$\hat{u}: OWNS$} \hspace{0 cm}
\\
\includegraphics[width=0.48\linewidth,trim={9cm 0cm 22cm 2cm},clip]{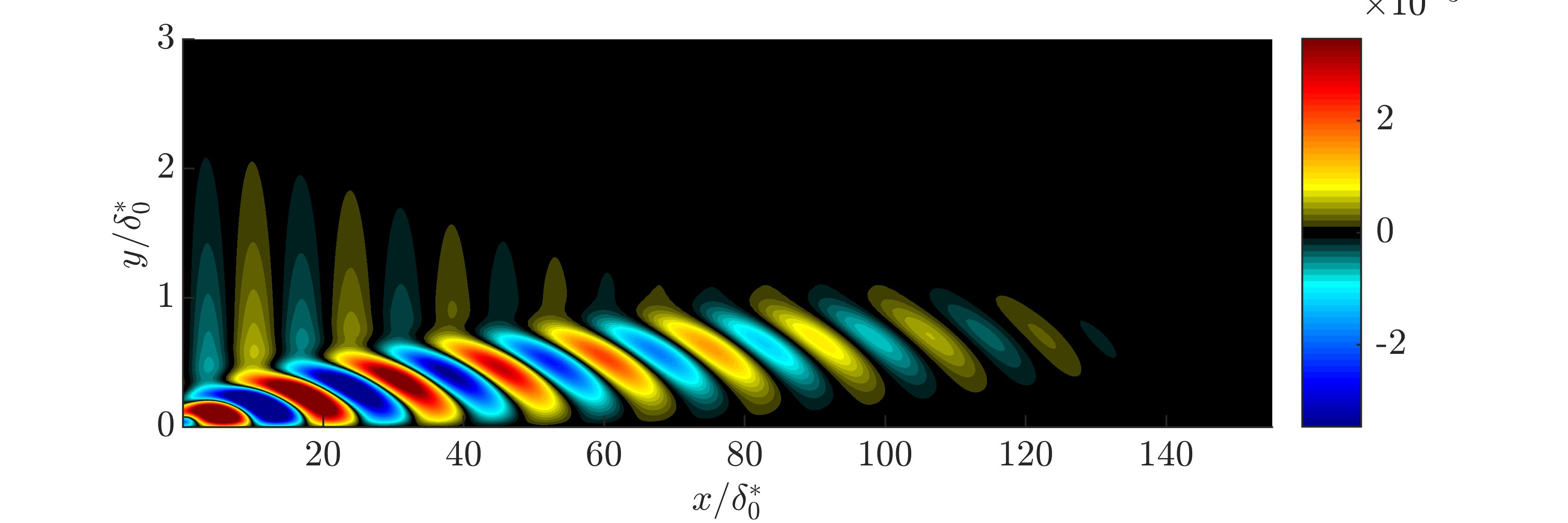}
\includegraphics[width=0.48\linewidth,trim={9cm 0cm 22cm 2cm},clip]{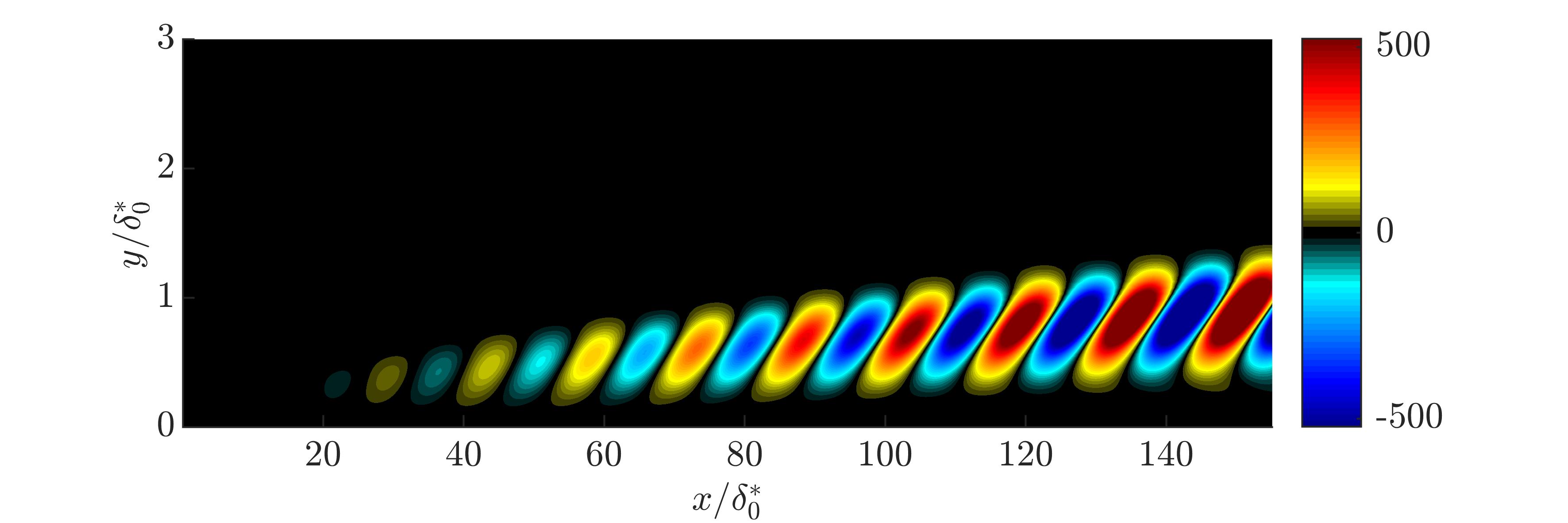}\\
\vspace{1cm}
 \includegraphics[width=0.4\linewidth,trim={0cm 0cm 0cm 0},clip]{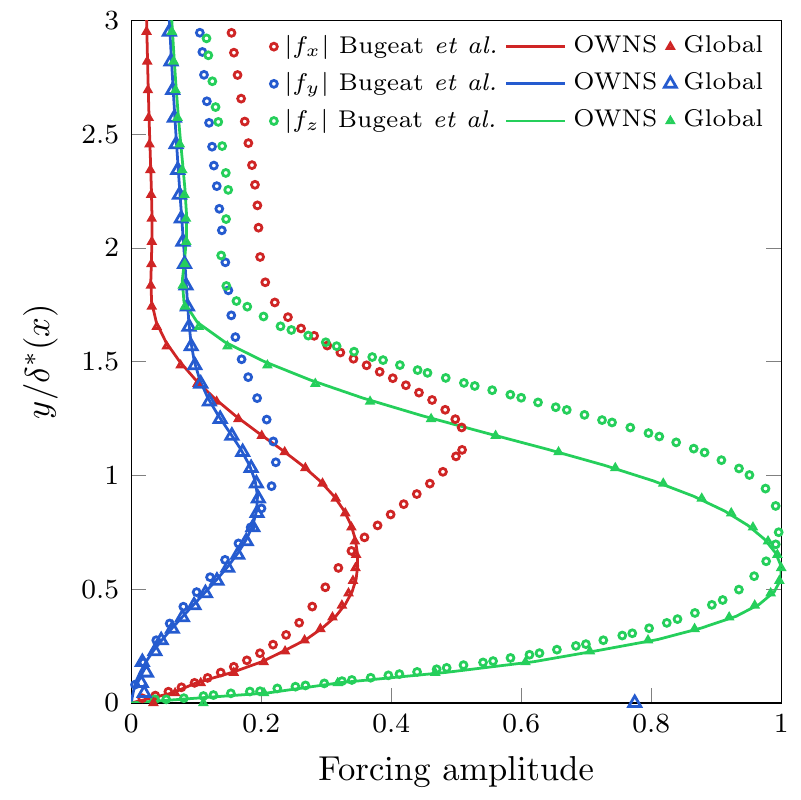}
 \hspace{1cm}
  \includegraphics[width=0.4\linewidth,trim={0cm 0cm 0cm 0},clip]{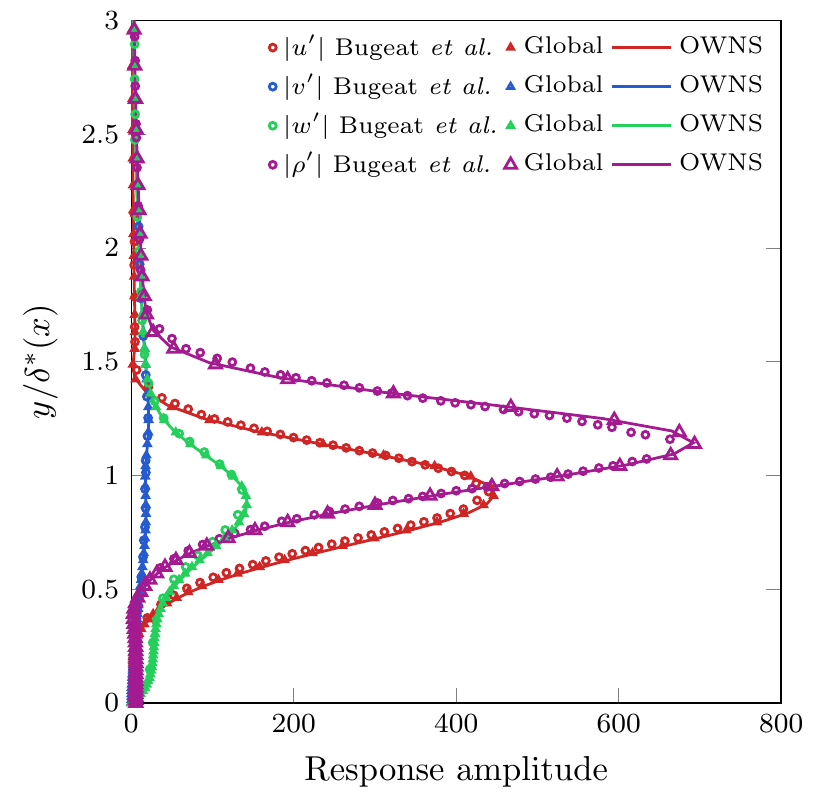}
 \end{center}
 \caption{Oblique first mode (unsteady 3D) optimal disturbances at $\omega=0.32$, $\beta=1.2$. Forcing and response amplitude components at $x/\delta^*_0=12$ (left) and $x/\delta^*_0=159$ (right). Circle symbols: \cite{bugeat20193d}; triangle symbols: global; solid lines: OWNS. }
 \label{fig:oblique}
\end{figure}


Contours of the optimal forcing and response for the oblique first mode are shown in figure \ref{fig:oblique} for the dominant components $f_w$ and $\hat{u}$ from the OWNS-P and global calculations. The optimal forcing field contains upstream-titled structures that are emblematic of the non-modal Orr mechanism. This generates an oblique wave response with relatively large streamwise velocity. As above, quantitative agreement is obtained between the global and OWNS-P results for all the perturbation input and output components. 

\begin{figure}
\begin{center}
\hspace{0cm} {$f_v: Global$} \hspace{4cm} {$u': Global$} \hspace{0 cm}  \\
\includegraphics[width=0.48\linewidth,trim={9cm 7cm 22cm 2cm},clip]{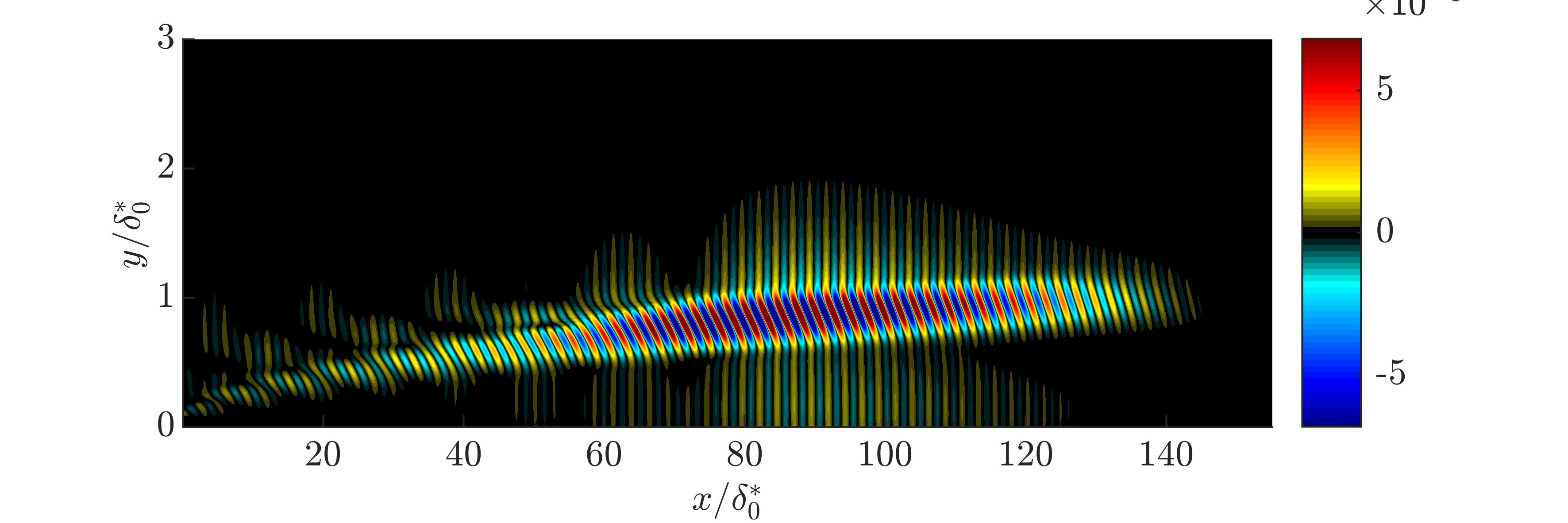}
\includegraphics[width=0.48\linewidth,trim={9cm 7cm 22cm 2cm},clip]{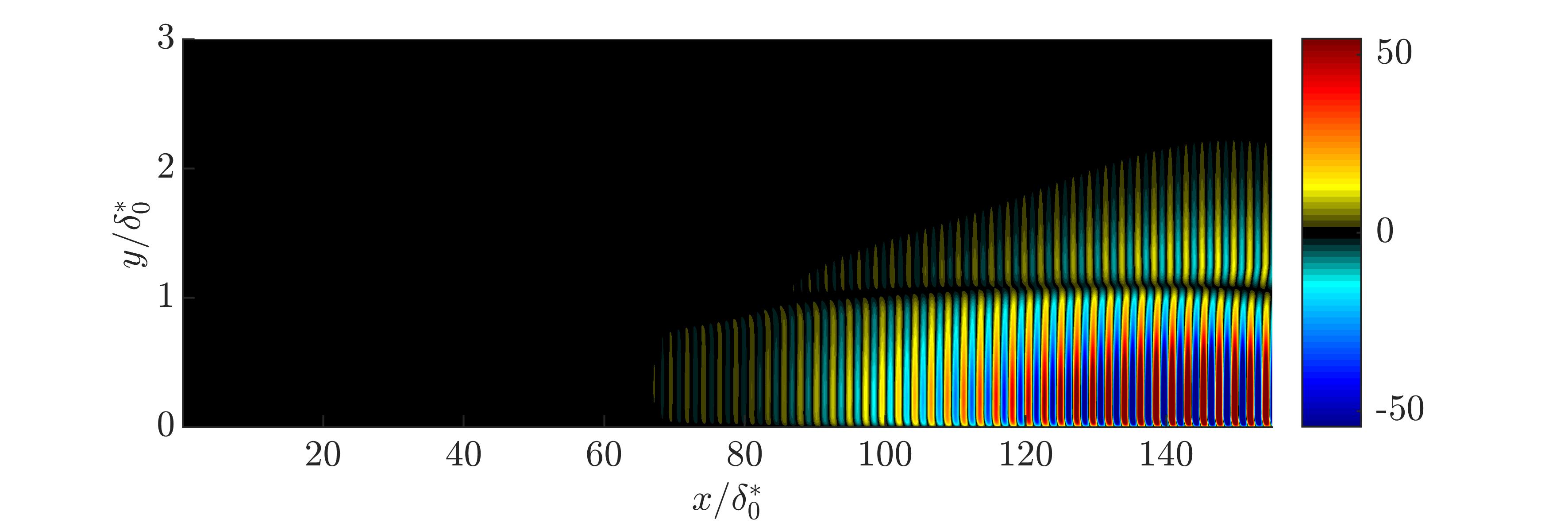}
\\
\hspace{0cm} {$f_v: OWNS$} \hspace{4cm} {$u': OWNS$} \hspace{0 cm}
\\
\includegraphics[width=0.48\linewidth,trim={9cm 0cm 22cm 2cm},clip]{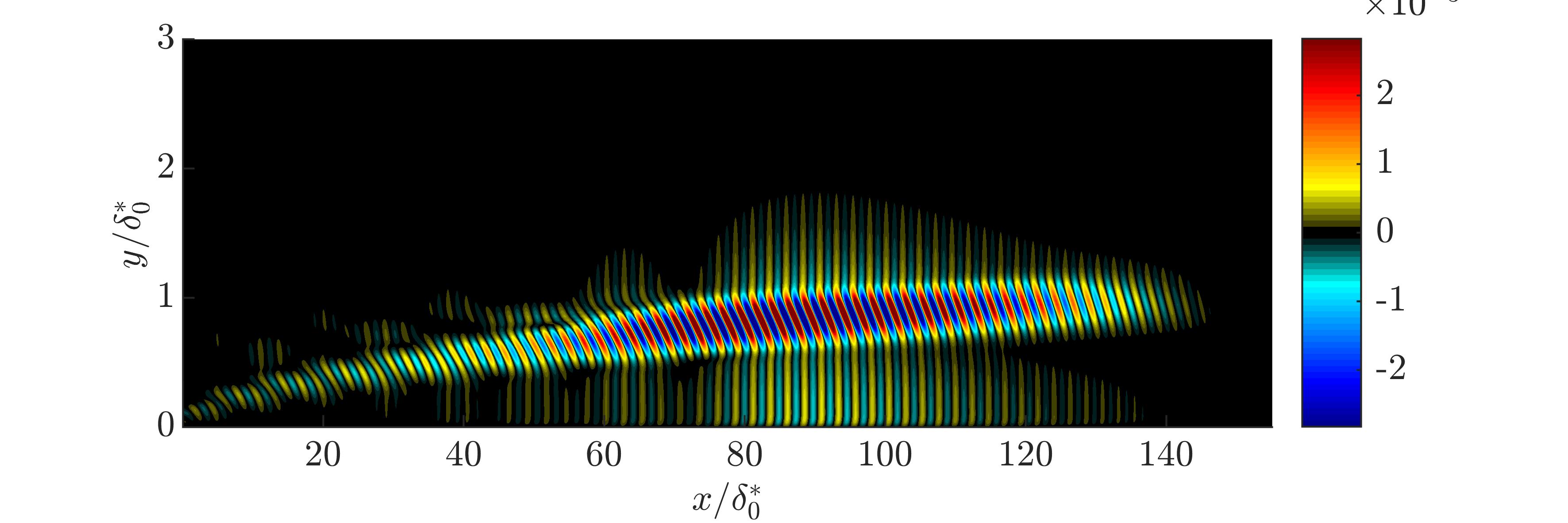}
\includegraphics[width=0.48\linewidth,trim={9cm 0cm 22cm 2cm},clip]{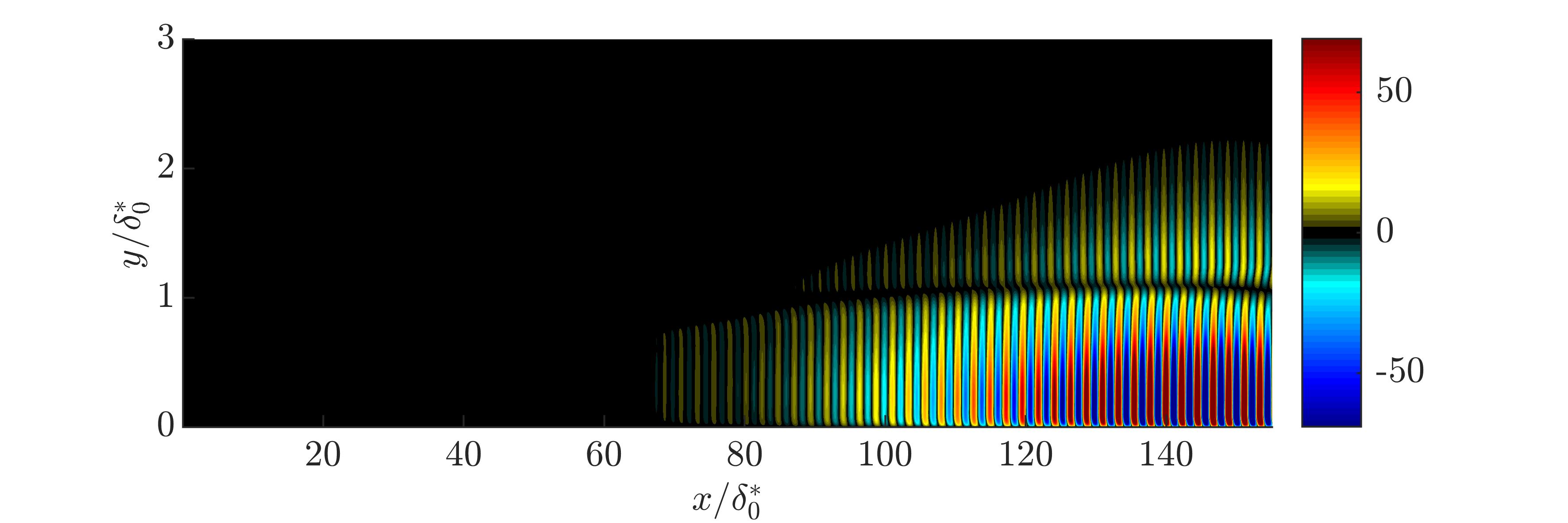}\\
\vspace{1cm}
 \includegraphics[width=0.4\linewidth,trim={0cm 0cm 0cm 0},clip]{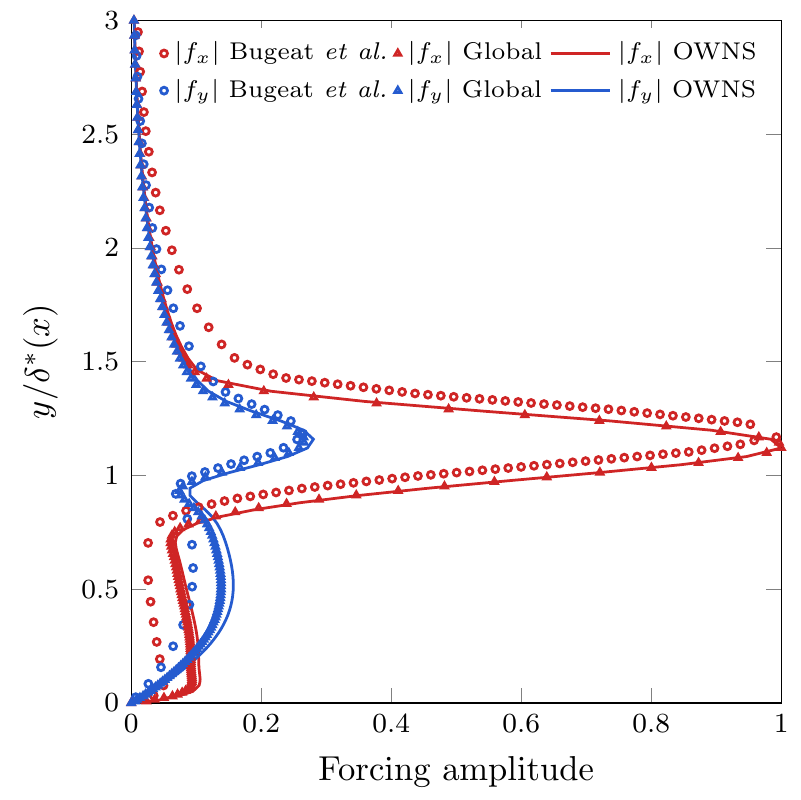}
 \hspace{1cm}
  \includegraphics[width=0.4\linewidth,trim={0cm 0cm 0cm 0},clip]{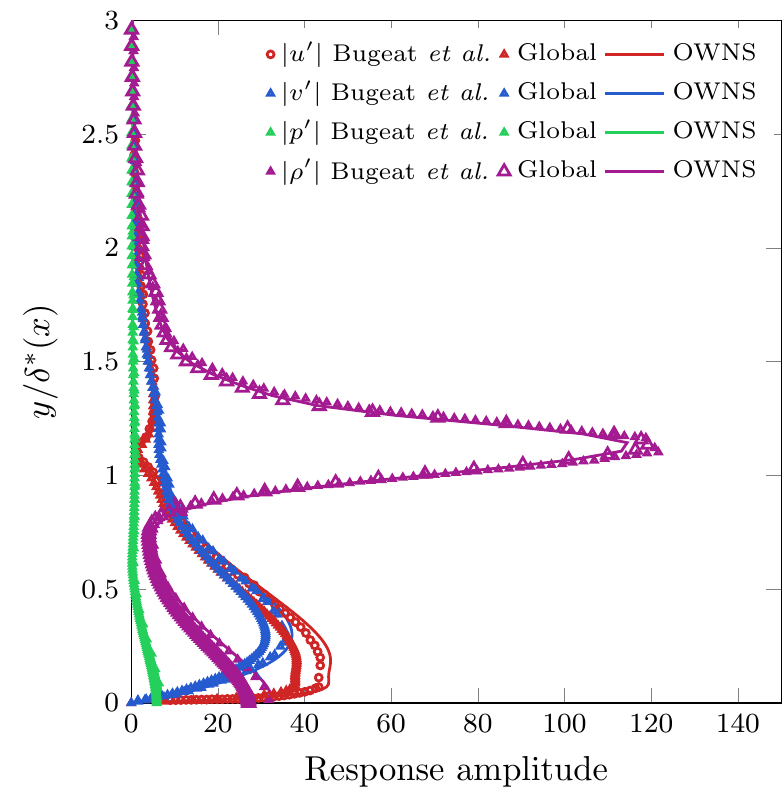}
  \end{center}
 \caption{Planar second mode (unsteady 2D) optimal disturbances at $\omega=2.5$, $\beta=0$. Forcing and response amplitude components at $x/\delta^*_0=90$ (left) and $x/\delta^*_0=148$ (right). Circle symbols: \cite{bugeat20193d}; triangle symbols: global; solid lines: OWNS.}
 \label{fig:secondmode}
\end{figure}

Lastly, in figure \ref{fig:secondmode} we compare the results for the second mode instability, following a similar procedure as in the two previous cases. We see the classical trapped acoustic waves between the wall and relative sonic line as well as thermodynamic amplification near the generalized inflection point in the response fields (two coexisting mechanisms). For such a response, the optimal forcing is localised near the generalized inflection point. Good agreement is also obtained between OWNS-P and our global results for this family of modes.

\section{Conclusions}\label{sec:conclusions}


In this paper, we have developed a computationally efficient approach for computing optimal frequency-response modes (known as resolvent or input/output modes) describing the fluctuating hydrodynamic and acoustic fields for flows with a slowly varying direction.  Specifically, we leverage the ability of the projection-based one-way Navier-Stokes (OWNS-P) equations, formulated in Part 1 of this two-part paper \citep{Towne2021fast}, to efficiently and accurately approximate the action of the resolvent operator on a forcing vector via well-posed spatial marching in the slowly varying direction.  Using an adjoint-based optimization framework, volumetric forcings that optimally excite and reveal the dominant three-dimensional instabilities of the flow are computed by marching the forward and adjoint OWNS-P equations in the downstream and upstream directions, respectively.  Thus, we bypass the solution of direct and adjoint globally-discretized Navier-Stokes equations, which can be computationally expensive or even intractable for three-dimensional inhomogeneous (base) flows. 

The optimal OWNS framework was demonstrated for a supersonic turbulent jet at Mach 1.5 and a transitional zero-pressure-gradient flat-plate boundary layer flow at Mach 4.5. We showed that three-dimensional resolvent modes can be obtained in a computational efficient way by spatial marching of the direct and adjoint OWNS-P equations in the slowly-varying streamwise direction. The framework was validated against global modes calculated by discretizing in all inhomogeneous spatial directions, which included multi-modal behaviour and interplay between various instability mechanisms, including lift-up, Orr, Kelvin-Helmholtz and acoustic modes. Excellent agreement between the results obtained from the global and OWNS-P methods demonstrate the ability of the latter to accurately retain downstream-traveling waves and eliminate upstream traveling waves, and also confirms the negligible influence of the upstream propagating waves for these problems.  While we considered two high-speed flow examples, the methodology is applicable to all flow regimes, provided the existence of a slowly varying direction.


The optimal OWNS framework presented here will enable calculations that reveal the dominant instability mechanisms of complex configurations that are presently intractable for global methods. Furthermore, the OWNS-P framework can be used for the calculation of nonlinear optimal disturbances \citep{rigas2021nonlinear} by including a finite number of harmonic balanced nonlinear interactions in the Lagrangian formulation. Calculation of nonlinear disturbances in compressible regimes has been attempted in the past using nonlinear PSE, which cannot account for multimodal instabilities or non-normal amplification mechanisms \citep{TowneTCFD2018}, which play a key role during the transition process.

\section*{Acknowledgments}

G.R., O.K. and T.C. acknowledge the support of the Boeing Company through a Strategic Research and Development Relationship Agreement CT-BA-GTA-1.  A.T. was funded in part by a catalyst grant from the Michigan Institute for Computational Discovery and Engineering (MICDE).  T.C. and O.K. also acknowledge recent support from the Office of Naval Research under grant N00014-21-1-2158.  O.K. also acknowledges a fellowship {fill in}.


\begin{appendices}

\section{Self-similar boundary layer solutions}\label{app:baseflow}

The two dimensional base flow is obtained by applying a Howarth-Dorodnitsyn transformation, under which the governing compressible boundary layer equations reduce to ODE's of the form \citep{Stewartson1964}
	\begin{subequations}
        \begin{align}
         & ff'' +  \left( \frac{\mu \rho}{\mu_\infty \rho_\infty} f'' \right)'  	= 0, \\
         & \left( \frac 1 {Pr} \frac{\mu \rho}{\mu_\infty \rho_\infty} T' \right)' + fT' +(\gamma-1) M_\infty^2   \frac{\mu \rho T_\infty}{\mu_\infty  \rho_\infty} f''^2	= 0,
        \end{align}
	\end{subequations}
where
$$
f' = f'(\eta) = U/U_\infty
$$
with corresponding boundary conditions
\begin{align*}
f(0)=f'(0)=0, \quad f'(\infty)=1, \quad T(\infty)=T_\infty, \quad T'(0)=0 ~\mathrm{(adiabatic)}.
\end{align*}
In the above, $()'$ quantities, denote derivatives with respect to the vertical self-similar $\eta$-direction.

\end{appendices}

\bibliographystyle{jfm}
\bibliography{jfmbib}

\end{document}